\documentclass[aps,prb,twocolumn,superscriptaddress,showpacs]{revtex4}
\usepackage{graphicx}
\usepackage{color}
\usepackage{amssymb}
\usepackage{amsmath}
\newcommand{\FG}[1]{Fig.~\ref{#1}}
\newcommand{\EQ}[1]{Eq.~(\ref{#1})}
\newcommand{\ea}{{\it et al.}}
\newcommand{\bm}[1]{\mbox{\boldmath$#1$}}
\pdfoutput=1
\begin{document}
\title{Interplay of growth mode and thermally induced spin 
accumulation in epitaxial 
Al/Co$_2$TiSi/Al and Al/Co$_2$TiGe/Al contacts}
\author{Benjamin Geisler}
\author{Peter Kratzer}
\author{Voicu Popescu}
\altaffiliation{E-mail: voicu.popescu@uni-due.de}
\affiliation{
Faculty of Physics and
Center for Nanointegration (CENIDE),
University of Duisburg-Essen,
Lotharstra{\ss}e\ 1, 47057 Duisburg, Germany}

\date{\today}

%%%%%%%%%%%%%%%%%%%%%%%%%%%%%%%%%%%%%%%%%%%%%%%%%%%%%%%%%%%%%%%%%%%%%%%%%%%
%%%%%%%%%%%%%%%%%%%%%%%%%%%%%%%%% ABSTRACT
%%%%%%%%%%%%%%%%%%%%%%%%%%%%%%%%%
\begin{abstract}
The feasibility of thermally driven spin injectors built
from half-metallic Heusler alloys inserted between aluminum
leads was investigated by means 
of {\em ab initio} calculations of the thermodynamic 
stability and electronic transport. 
We have focused on two main issues
and found that: (i)  the interface
between Al and the 
closely lattice-matched Heusler
alloys of type Co$_2$Ti$Z$ ($Z=$ Si or Ge)
is stable under various growth conditions; and
(ii) the conventional and spin-dependent Seebeck coefficients
in such heterojunctions exhibit a strong dependence on
both the spacer and the atomic composition of the
Al/Heusler interface. The latter quantity gives a measure
of the spin accumulation and varies between $+8$~$\mu$V/K
and $-3$~$\mu$V/K near $300$~K,
depending on whether a Ti-Ge or a Co-Co
plane makes the contact between Al and Co$_2$TiGe
in the trilayer. Our results show that it is in 
principle possible to tailor the spincaloric effects 
by a targeted growth control of the samples.
\end{abstract}
%%%%%%%%%%%%%%%%%%%%%%%%%%%%%%%%%
%%%%%%%%%%%%%%%%%%%%%%%%%%%%%%%%%%%%%%%%%%%%%%%%%%%%%%%%%%%%%%%%%%%%%%%%%%%
%%%%%%%%%%%%%%%%%%%%%%%%%%%%%%%%%%%%%%%%%%%%%%%%%%%%%%%%%%%%%%%%%%%%%%%%%%%
% 72.10.-d - Theory of electronic transport; scattering mechanisms
% 72.15.Jf - Thermoelectric and thermomagnetic effects
% 73.50.Jt - Galvanomagnetic and other magnetotransport effects 
%            (including thermomagnetic effects) 
%%%%%%%%
%%%%%%%%%%%%%%%%%%%%%%%%%%%%%%%%%
\maketitle
%%%%%%%%%%%%%%%%%%%%%%%%%%%%%%%%%
%%%%%%%%%%%%%%%%%%%%%%%%%%%%%%%%%%%%%%%%%%%%%%%%%%%%%%%%%%%%%%%%%%%%%%%%%%
%SSSSSSSSSSSSSSSSSSSSSSSSSSSSSSSSSSSSSSSSSSSSSSSSSSSSSSSSSSSSSSSSSSSSSSSSSSSSSSS
%SSSSS
%SSSSS
\section{Introduction}\label{SecIntro}

A central topic of spintronics is the design and realization of
spin injectors.\cite{ZFDS04} These are contacts that allow one 
to induce a spin accumulation (in general terms, an imbalance 
of the chemical potential for the two spin channels) 
in a substrate. While spin injectors integrated within
the technological Si standard\cite{Jansen:12} would
be most desirable, their actual implementation is hindered by various
factors. 

Indeed, while the spin injection may in principle be 
accomplished using magnetic transition metals and their alloys,
the preparation of an atomically well-defined interface between 
a transition metal and the Si surface is very difficult due 
to the high tendency of exothermic silicide 
formation.\cite{BSR01,KBH+03,WKS05,GKS+12,GK13}
As an alternative one could use a light $sp$~metal such as 
Al as a contact layer between Si and the ferromagnet. 
Ohmic contacts between Si and Al are well-studied and form a standard 
component in Si device technology.\cite{Car76}
More importantly, Al displays a very large spin 
diffusion length,\cite{Fabian99} and hence Al leads are suitable to 
conduct a spin-polarized current without substantial losses. 
The problem is then rendered into finding an appropriately matching 
ferromagnetic system acting as spin injector.

For this task, the ferromagnetic Heusler alloys have been 
recently investigated theoretically as promising
candidates.\cite{CGC+11}
In their ideal $L2_1$ crystal structure 
these systems are ferromagnetic half-metals,\cite{KFF07}
which in principle allows a high degree of spin polarization
of the injected current.  
Moreover [as shown in \FG{FIGStrucs}(a)] the $L2_1$ structure
can be matched by an $A1$ structure rotated by $45^\circ$ 
about the $z$-axis. 
In particular, the Co-based full Heusler alloys
Co$_2$TiSi or Co$_2$TiGe recommend themselves for integrated 
spin injectors in combination with an Al contact layer
as the experimentally determined lattice mismatch is small
(less than $\simeq 2$~\%).

Applying an external voltage to a Heusler/Al/Si junction may
not be, however, the best way to achieve a high value of spin 
accumulation. A considerable obstacle in the practical 
realization of a metal-semiconductor spin injector is the so-called 
conductivity mismatch\cite{Schmidt2000} between the different 
materials. This leads to a potential drop in locations where it is not 
useful for the device, while the spin accumulation in the
semiconductor itself might remain small. 
It has been suggested\cite{SBAvW10} that this
difficulty could be overcome by applying an external temperature
gradient, rather than an external voltage, in order to induce the spin
accumulation: Exploiting the Seebeck effect, a temperature gradient
across the contact directly results in a difference in the chemical
potentials in the two spin channels due to the spin-dependence of the
Seebeck coefficient. This difference is independent of an injected
current and hence unaffected by a possible mismatch in conductivity. 
Experimental studies\cite{BOG+10,BFB+10} on bulk samples have shown that
the Ti-based Heusler alloys, such as Co$_2$TiSi or Co$_2$TiGe, 
display large Seebeck voltages (tens of $\mu$V/K)
of negative sign under an applied temperature gradient. 
One would expect from these findings  
that these materials,
in conjunction with their half-metallic electronic structure, 
might also show a large {\em spin}-dependence of the Seebeck 
coefficient, thus making them suitable candidates for 
spin injectors based on spincaloric effects.
Bulk Al, on the other hand, is known to exhibit
a rather small, negative Seebeck coefficient: 
$-1.78$~$\mu$V/K at room temperature, and somewhat
increased at lower temperatures due to phonon drag.\cite{Hue68,BA99}
Thus, using Al as contact material will have little impact on the
thermoelectric properties of the junction.

It is the aim of our present investigations to assess
the ability of Al/Co$_2$Ti$Z$/Al
($Z$ = Si or Ge) trilayers to serve 
as thermally driven spin injectors. This has been 
accomplished
by performing first-principles calculations of the 
electronic structure and of the thermoelectric 
transport properties. While ignoring the other side
of the device, the spin injector-semiconductor contact,
we focused on two specific problems for the
Al/Co$_2$Ti$Z$/Al systems: (i) the stability and the electronic
structure of the Al/Heusler interface,
and (ii) whether they open a promising path towards
maximizing the thermally induced spin accumulation.
We further provide a detailed insight in the electronic 
transport mechanisms in these junctions, accounting for a 
realistic morphology of the interface and focusing on the electronic
structure contribution to the transmission and Seebeck coefficient.

Our results show that the formation energy of the Al/Heusler
interface is negative, which means that these interfaces are
stable. While both Heusler alloys can match the Al
substrate either with a Ti-$Z$ or a Co-Co atomic plane,
the former needs non-equilibrium growth conditions
to avoid the formation of competing Ti$Z$ compounds. 
In the Al/Co$_2$Ti$Z$/Al trilayers both conventional and 
spin-dependent Seebeck effects are found to be sensitive 
to the specific atomic structure of the Al/Heusler interface
and the actual spacer material. We compare the results obtained for
the heterostructures with those of the bulk Heusler alloys
in their cubic and tetragonally distorted structures. We find
that a subtle interplay between biaxial strain and the 
transmission channel selectivity governs
the transport properties of the investigated trilayer
systems. In particular, for a thin Co$_2$TiSi or Co$_2$TiGe layer 
terminated by a Ti-Si or Ti-Ge plane,
the spin-dependent Seebeck coefficient is 
positive and of the same order of magnitude as the conventional, 
spin-averaged Seebeck coefficient. For a Co-Co-terminated
Al/Co$_2$TiGe/Al trilayer both coefficients are negative.
This suggests the possibility of achieving 
a large and stable spin accumulation by 
employing appropriate growth conditions
during sample preparation.

%SSSSSSSSSSSSSSSSSSSSSSSSSSSSSSSSSSSSSSSSSSSSSSSSSSSSSSSSSSSSSSSSSSSSSSSSSSSSSSS
%SSSSS
%SSSSS
\section{Description of structures and applied methods}
\label{SecMethod}

\subsection{Setting up the 
               Al/Heusler/Al trilayer system}\label{SecGeom}

At the center of our investigations are the Al/Co$_2$Ti$Z$/Al
junctions in trilayer geometry, with $Z=$ Si or Ge. 
We give here the geometric
arguments for the feasibility of such heterostructures and
will show later by first-principles calculations that they
are energetically stable. 
The Co$_2$Ti$Z$ compounds belong to the class of 
full Heusler alloys\cite{GFP11} of type
$X_2YZ$ which crystallize in the cubic $L2_1$ 
structure. This crystal structure, depicted in 
\FG{FIGStrucs}(a), has a face-centered-cubic (fcc)
primitive cell with four inequivalent atomic sites. 
It is usually described as  
four inter-penetrating fcc sublattices, respectively 
occupied by the 
$X$, $Y$, $X$, and $Z$ atoms,
shifted against each other by $(a/4,a/4,a/4)$,
with $a$ being the cubic lattice constant. Alternatively,
one can represent the $L2_1$ structure by 
two inter-penetrating $XY$ and $XZ$
zinc-blende structures, with the latter shifted by $(0,0,a/2)$.
Aluminum, on the other hand, possesses the simple
$A1$ crystal structure, consisting 
of an fcc Bravais lattice with one atom per unit cell.
As illustrated in \FG{FIGStrucs}(a), the 
$A1$ structure, rotated by $45^\circ$ about the $z$ axis,
represents a natural continuation of the $L2_1$ structure,
enabling a perfect epitaxial match of the two systems 
if their lattice constants are in a ratio of
$a(L2_1)/a(A1)=\sqrt{2}$. The two Heusler alloys chosen for the
present study deviate only slightly from this condition, namely 
by $1.07$~\% for Co$_2$TiSi/Al and $2.70$~\% 
for Co$_2$TiGe/Al.

%FFFFFFFFFFFFFFFFFFFFFFFFFFFFFFFFFFFFFFFFFFFFFFFFFFFFFFFFFFFFFFFFFFFFFFFFFFFFF
%FFFFFFFFFFFFFFFFFFFFFFFFFFFFFFFFFFFFFFFFFFFFFFFFFFFFFFFFFFFFFFFFFFFFFFFFFFFFF
%FFFFFFF
\begin{figure*}
    \centering
  \includegraphics[width=\textwidth]{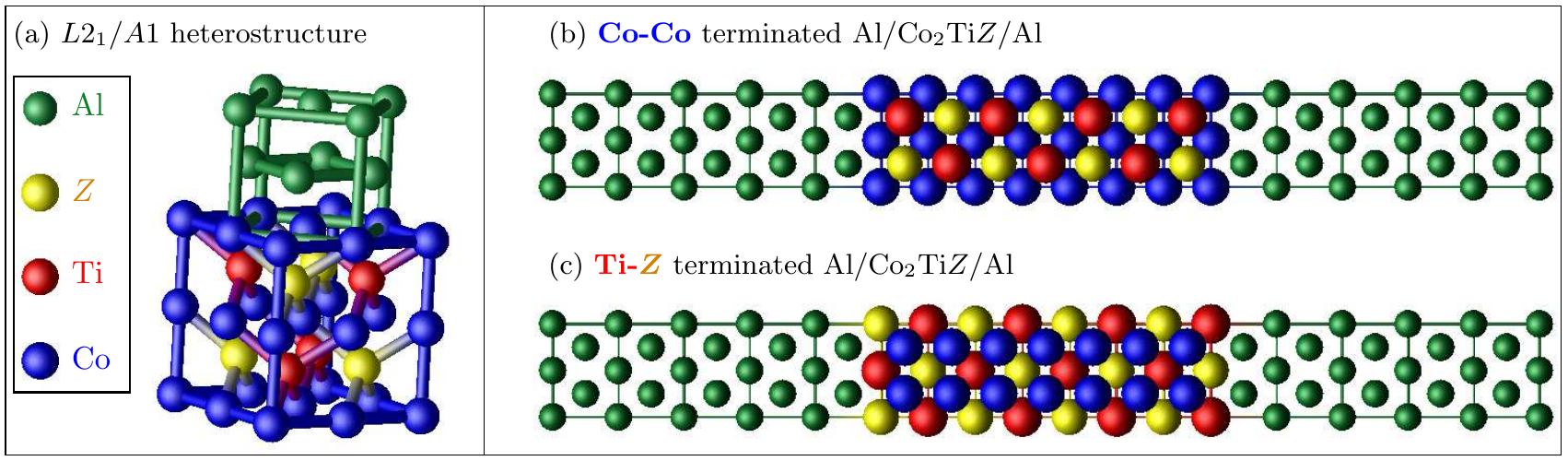}
  \caption{(Color online) (a) Structural model of the
   Co$_2$Ti$Z$ ($Z$=Si/Ge) full Heusler $L2_1$ structure matched to
   the $A1$ structure rotated by an angle of $45^\circ$ 
   about the vertical ($z$) axis. Panels (b) and (c) show $(110)$-projected 
   views of the
   tetragonal Al/Co$_2$Ti$Z$/Al supercells used in our calculations 
   with either Co-Co or Ti-$Z$ 
   termination of the Al/Heusler interface.}
  \label{FIGStrucs}
\end{figure*}
%FFFFFFF
%FFFFFFFFFFFFFFFFFFFFFFFFFFFFFFFFFFFFFFFFFFFFFFFFFFFFFFFFFFFFFFFFFFFFFFFFFFFFF
%FFFFFFFFFFFFFFFFFFFFFFFFFFFFFFFFFFFFFFFFFFFFFFFFFFFFFFFFFFFFFFFFFFFFFFFFFFFFF

Bringing together the two structures, on the other hand, requires the
use of a tetragonal unit cell of in-plane lattice constant 
$a_{\rm tet}=a(A1)$, with unit vectors oriented along the 
$L2_1$-$(110)$ and $(\bar{1}10)$ directions, and
of varying length $c$ along the common $(001)$ direction. Each
plane of the tetragonal unit cell contains two inequivalent atomic
sites. The Al/Co$_2$Ti$Z$ heterostructures were modeled 
with such supercells with $a_{\rm tet}$ fixed
to the equilibrium lattice constant of Al metal.
We considered two terminations of the Heusler materials
at the interface to the Al electrodes:
a pure Co-Co layer and a mixed Ti-$Z$ layer, 
where $Z=$~Si or Ge, as illustrated in \FG{FIGStrucs}(b) and (c).
The supercells contain $2\times 10$ Al planes and
$8+7$ Co (Ti-$Z$) and Ti-$Z$ (Co) Heusler planes, leading
to an appropriate, well-converged potential in the Al electrodes.
We constructed the supercells such that the Heusler atoms
simply continue  the Al fcc lattice by occupying the hollow sites
next to the interface.

All internal atomic positions have been accurately optimized using
Hellmann-Feynman forces to reduce the force components below
$1$~mRy/bohr and the energy changes below $0.1$~mRy.
Moreover, the length of the tetragonal Al/Heusler/Al supercells
has been optimized for each considered structure in order to determine
the ideal, energy-minimizing Al-Heusler spacing.

% Similar tetragonal unit cells (with and without epitaxial strain)
% have been employed for transport calculations of bulk Heusler materials.

\subsection{Electronic structure calculations}

The electronic structure and transport calculations 
have been performed within the framework of spin-polarized 
density functional theory (DFT)
employing the plane-wave pseudopotential method 
as implemented in the Quantum Espresso code,\cite{PWSCF}
with the PBE\cite{PeBu96}
generalized gradient approximation (GGA)
parametrization of the exchange-correlation 
functional. 
Wave functions and density have been expanded into plane waves
up to cutoff energies of $40$~Ry and $400$~Ry, respectively.
The neighborhood of atom centers has been approximated by
self-created ultrasoft pseudopotentials (USPPs),\cite{Vanderbilt:1990}
treating the atomic Si $3s$, $3p$,
Ge $3d$, $4s$, $4p$, Co $3d$, $4s$, $4p$,
and Ti $3s$, $3p$, $3d$, $4s$, $4p$ subshells
as valence states.\cite{AlNote}
For Si, Ge, and Co a non-linear core correction \cite{LoFr82} was included.
In the pseudopotential creation process a 
scalar relativistic approximation was applied to
the electron motion. 
A Methfessel-Paxton smearing \cite{MePa89} of 10~mRy
has been applied to the Brillouin zone (BZ) sampling 
performed with different Monkhorst-Pack $k$-point grids:\cite{MoPa76}
For the Heusler fcc bulk calculations, 
we used a $16 \times 16 \times 16$ grid;
$20 \times 20 \times 14$ for the tetragonal Heusler
cells and $20 \times 20 \times 20$ for the Al 
cubic cell. Finally, for the 
Al/Heusler/Al supercells the $k$-point grid was 
$16 \times 16 \times 4$.
All grids were chosen in such a way
that they did not include the $\Gamma$ point
and deliver accurately converged Fermi energies and
potentials. Post-processing of densities of states was 
performed with much denser
$k$-point grids that included the $\Gamma$ point.

\subsection{Interface formation energies}\label{SubsecIFACE}

Based on the results obtained for
the constructed supercells it is possible,
using an {\em ab initio} thermodynamic
approach,\cite{RSS04} to derive interface energies,
\begin{equation}
\label{IF-Energies}
\begin{aligned}
\gamma(\mu_{\text{Co}})  = & \frac {1}{2A}
                         \left[
                           E_{\text{sc}}
                           - N_{\text{Al}} E_{\text{Al}}
                           - \mu_{\text{Co}} (N_{\text{Co}}- 2 N_{Z}) \right.\\
         &\left.\quad\quad - N_{Z} E_{\text{Co$_2$Ti$Z$}}\right]
         \enspace,
\end{aligned}
\end{equation}
which provide information about the stability of different
interfaces in thermodynamic equilibrium.
Here $E_{\rm sc}$, $E_{\rm Al}$, and $E_{\text{Co$_2$Ti$Z$}}$ are
the DFT total energies of the considered supercell, Al bulk, and the
respective Heusler bulk; $A$ is the interface area and 
the $N_{i}$'s denote the different numbers of atoms of 
species $i$ in the supercell.
Corrections to the interface energy as they arise, for instance,
due to the phonon free energy at finite temperatures are neglected,
since they will certainly be similar for the different structures.
In writing down this expression, one 
considers Co, Ti, and $Z$ to be in equilibrium with the
bulk phase of Co$_2$Ti$Z$:
\begin{equation}  \label{cotizbulk}
  E_{{\rm Co_2Ti}Z}  = 2\,\mu_{\rm Co} + \mu_{\rm Ti} + \mu_Z\enspace. 
\end{equation}
In addition, for the supercells considered here,
the interface energy explicitly depends only 
on $\mu_{\text{Co}}$, since in all cases $N_{\text{Ti}} = N_{Z}$. 
A reduced chemical potential
$\tilde\mu_{\text{Co}} = \mu_{\text{Co}} - E_{\text{Co}}$
can be defined using the DFT energy of hcp Co, $E_{\text{Co}}$.
Since this quantity corresponds to the formation of the 
Co metal, the inequality
\begin{equation}
  \label{MuUpBound}
\tilde\mu_{\text{Co}} \leq 0
\end{equation}
can be seen as an upper bound of the reduced chemical potential
of Co. Different lower bounds can be determined by taking
into account the formation of competing
compounds such as CoTi, Co$Z$, and Ti$Z$. Assuming, for example,
Co and Ti to be in equilibrium with the CoTi bulk phase,
\begin{equation}
  E_{\rm CoTi}  = \mu_{\rm Co} + \mu_{\rm Ti}\label{cotibulk}\enspace,
\end{equation}
it follows, using \EQ{cotizbulk},
\begin{equation}\label{mucolow1}
  E_{{\rm Co_2Ti}Z} - E_{\rm CoTi}  \leq \mu_{\rm Co} + \mu_Z
\end{equation}
or
\begin{equation}\label{mucolow}
\tilde\mu_{\text{Co}} \geq - \tilde\mu_Z + 
     \left[ E_{{\rm Co_2Ti}Z} - E_{\rm CoTi} - 
            E_{\text{Co}}- E_Z \right] \enspace,
\end{equation}
where, except for the parameter 
$\tilde\mu_Z=\mu_Z-E_Z$,
all quantities on the right-hand side are directly accessible from
DFT calculations for the corresponding systems. Other
lower boundaries for $\tilde\mu_{\text{Co}}$ can be determined
analogously.

\subsection{Calculation of Seebeck coefficients}

For the transport properties we considered an open quantum
system consisting of a scattering region comprising the Heusler material
and interfaces to the electrode material, and 
the left and right semi-infinite electrodes.
From the accurately converged DFT potentials of the leads and the
scattering regions,
transport coefficients have been calculated separately for
both spin channels using a method following
Refs.~\onlinecite{SmogunovTosatti:04}
and~\onlinecite{ChoiIhm:99}.
In order to sample the two-dimensional (2D) BZ
on a reasonable computational time scale, we have 
massively parallelized the method. 
Convergence of the energy- and spin-resolved
transmission probability~${\cal T}_\sigma(E)$,
\begin{equation}
  \label{TofE}
  {\cal T}_\sigma(E) = \frac{1}{A_{\rm BZ}} \int {\rm d}^2 k_\parallel
  \,
  {\cal T}_\sigma(\vec k_\parallel,E)
\text{,}
\end{equation}
with respect to the
$k_\parallel$-point grid has been found to be attained with
a $201 \times 201$ $k_\parallel$-points regular mesh.
The Seebeck coefficients have been evaluated using
the approach of Sivan and Imry\cite{SI86}
starting from the central quantity ${\cal T}_\sigma(E)$
and the Fermi occupation function $f_0(E,T,\mu)$. 
Within Mott's two current model
the spin-projected conductance is expressed as
\begin{equation}
  \label{Conduct}
  G_\sigma(T)  = -\frac{e^2}{h} \,
        \int {\rm d}E \, \frac{\partial f_0}{\partial E} \,
        {\cal T}_\sigma(E)\enspace,
\end{equation}
while the spin-projected Seebeck coefficient takes on the 
form 
\begin{equation}
  \label{Seebeck}
  S_\sigma(T)  = -\frac{1}{eT} \,
        \frac{{\displaystyle \int {\rm d}E \,
           \frac{\partial f_0}{\partial E}
               (E-E_{\rm F}) \, {\cal T}_\sigma(E)}}
             {{\displaystyle\int {\rm d}E \,
         \frac{\partial f_0}{\partial E} \, {\cal T}_\sigma(E)}}\enspace.
\end{equation}
Using the two quantities above the effective (also called charge)
Seebeck coefficient can be expressed as
\begin{equation}
  \label{EffSeeb}
  S_{\rm eff} = \frac{G_\uparrow\,S_\uparrow +
                         G_\downarrow\,S_\downarrow}
                    {G_\uparrow + G_\downarrow}\enspace,
\end{equation}
and the spin-dependent Seebeck coefficient by the
formula
\begin{equation}
  \label{SpinSeeb}
  S_{\rm spin} = \frac{G_\uparrow\,S_\uparrow -
                         G_\downarrow\,S_\downarrow}
                    {G_\uparrow + G_\downarrow}\enspace,
\end{equation}
with the temperature argument $T$ omitted.
The energy integration in Eqs.~\eqref{Conduct} and~\eqref{Seebeck}
consists of two steps: An explicit calculation of ${\cal T}_\sigma(E)$ 
on a regular $E$ mesh with a $15$~meV spacing followed by an 
interpolation of ${\cal T}_\sigma(E)$ on a refined $E$ mesh with 
a $1.36$~meV ($0.1$~mRy) spacing. 

The formalism adopted here to calculate the transport properties
only considers elastic scattering of 
the electrons by the interfaces, whereas inelastic scattering
processes, e.g., by phonons or spin fluctuations, are
neglected. These scattering mechanisms 
are expected to become more and more important
at high temperatures.
For example, including the electron scattering by spin
fluctuations in various ferromagnetic metals and alloys
had lead to a better agreement of the 
temperature dependent resistivity with experimental 
data.\cite{WSvsB09,KDT+12} A recent study\cite{KMWB14} 
investigated the effect of spin disorder on the magneto-thermoelectric 
phenomena in nano-structured Co systems. It could be 
demonstrated that, while the spin-dependent electron 
scattering does indeed influence the spin-caloric transport 
coefficients at high temperatures, this influence is strongly
case dependent, such that, without an explicit calculation,
no actual quantitative or qualitative predictions can be made.
In the following, we restrict our investigations to a temperature range 
below $350$~K and assume that the trends determined
on the basis of the electronic contribution alone are relatively 
reliable. We also neglect the possible magnetic anisotropy 
of the Seebeck coefficient\cite{PK13,WKE14} which may be 
induced by the spin-orbit coupling, since the elemental constituents 
of the investigated systems are relatively light.

%SSSSSSSSSSSSSSSSSSSSSSSSSSSSSSSSSSSSSSSSSSSSSSSSSSSSSSSSSSSSSSSSSSSSSSSSSSSSSSS
%SSSSS
%SSSSS
\section{Ground state properties of bulk
Co\bm{_2}TiSi and 
Co\bm{_2}TiGe}\label{SecResGSbulk}

The electronic, magnetic, and transport 
properties of the {\em bulk} Co$_2$Ti$Z$ Heusler alloys (with $Z$ being a
group IV element) are well documented in
the literature.\cite{LLBS05,KFF07,SSK10,BFB+10} 
We refer the reader in particular 
to the systematic study presented by
Barth \ea\cite{BFB+10} which combines full-potential LAPW-based
theoretical investigations with various experimental observations and
provides an exhaustive overview of the various properties
of the Co$_2$Ti$Z$ system. We use their all-electron results to 
assess the quality of our pseudopotential approach.

In addition, this section will present results obtained for 
the tetragonally distorted Heusler systems 
Co$_2$TiSi and Co$_2$TiGe in the 
so called free standing epitaxial geometry.
This concept designates
a partially constrained configuration in which the in-plane lattice
constant is fixed, typically to that of a substrate. 
Along the perpendicular direction, both 
atomic and unit cell parameter relaxation is allowed to occur 
in order to minimize the total energy. In such a setup, no actual
interaction with the substrate is accounted for,
isolating this way the effect of the epitaxial biaxial strain on 
the electronic structure of the material under investigation.
These results will be used later as reference for the actual 
Al/Heusler/Al trilayers.

%sssssssssssssssssssssssssssssssssssssssssssssssssssssssssssssssssssssssssssssss
%sssss
%sssss
\subsection{Results for the cubic \bm{L2_1} phase}\label{SubsecL21}

Table~\ref{TabGSabEQ} gives the
results of total energy minimizations for 
Co$_2$TiSi and Co$_2$TiGe with the equilibrium lattice constant $a$ 
and bulk modulus $B$ obtained from a fit to the Birch-Murnaghan 
equation of state.  
A comparison with the theoretical and experimental literature 
data evidences the excellent agreement with previous GGA-PBE 
results which are, in turn, only a few percent off 
the experimentally determined lattice constants.

%TTTTTTTTTTTTTTTTTTTTTTTTTTTTTTTTTTTTTTTTTTTTTTTTTTTTTTTTTTTTTTTTTTTTTTTTTTTTT
%TTTTTTTTTTTTTTTTTTTTTTTTTTTTTTTTTTTTTTTTTTTTTTTTTTTTTTTTTTTTTTTTTTTTTTTTTTTTT
%TTTTTTT
\begin{table}[ht]
\begin{tabular}{lrrrl}
\hline\hline
System & \multicolumn{2}{c}{Lattice constants} & 
         \multicolumn{1}{c}{$B$} & Source \\
       & \multicolumn{1}{c}{$a$ (\AA)} & 
         \multicolumn{1}{c}{$c$ (\AA)} & 
         \multicolumn{1}{c}{(GPa)}  & \\\hline\hline
Co$_2$TiSi $L2_1$
  & 5.756 &  &  204 & present work\\\hline
  & 5.753 &  &  210 & Theory\cite{LLBS05} \\
  & 5.758 &  &  207 & Theory\cite{BFB+10} \\
  &  5.74 &  &      & Experiment\cite{WZ73,CPAS93}\\
  & 5.849 &  &      & Experiment (300~K)\cite{BFB+10}\\ \hline\hline
Co$_2$TiGe $L2_1$
  & 5.848 &  & 189  & present work\\\hline
  & 5.842 &  & 193 & Theory\cite{LLBS05}\\
  & 5.850 &  & 190 & Theory\cite{BFB+10} \\
  & 5.83(1) &  &   &Experiment\cite{WZ73,CPAS93}\\
  & 5.820   &  &   &Experiment (300~K)\cite{BFB+10}\\\hline\hline
Co$_2$TiSi epi 
  & 5.695 & 5.819 & 290 & present work\\\hline\hline
Co$_2$TiGe epi 
  & 5.695 & 6.056 & 234 & present work\\\hline\hline
\end{tabular}
\caption{Calculated ground-state equilibrium lattice constants and
  bulk moduli of the Heusler alloys 
  Co$_2$TiSi and Co$_2$TiGe. The results are given for
  the cubic $L2_1$ phase as well as for the tetragonally
  distorted structure with a fixed in-plane lattice constant
  $a(L2_1)=a_0\sqrt{2}$, which corresponds to
  Heusler alloys epitaxially grown on fcc-Al(001) of
  lattice constant $a_0=4.027$~\AA.
  The results for the cubic systems are compared with
  available theoretical and experimental literature 
  data.}\label{TabGSabEQ}
\end{table}
%TTTTTTT
%TTTTTTTTTTTTTTTTTTTTTTTTTTTTTTTTTTTTTTTTTTTTTTTTTTTTTTTTTTTTTTTTTTTTTTTTTTTTT
%TTTTTTTTTTTTTTTTTTTTTTTTTTTTTTTTTTTTTTTTTTTTTTTTTTTTTTTTTTTTTTTTTTTTTTTTTTTTT

In their paper, Barth \ea\cite{BFB+10} characterize the Co$_2$Ti$Z$
compounds as itinerant ferromagnetic 
half-metals and contrast their behavior to that
of other Co-based Heusler alloys that have late $3d$ metals, 
e.g., Mn or Fe, on the $Y$ position. In addition, the samples 
prepared are reported to have nearly ideal $2:1:1$ stoichiometry and
magnetic moments of
$1.96$~$\mu_{\rm B}$ for Co$_2$TiSi and 
$1.94$~$\mu_{\rm B}$ for Co$_2$TiGe.
These combined results appear to hint towards the
half-metallicity of these systems as well as the absence of
substitutional (also called native) disorder 
between the four fcc sublattices that has been shown to
influence significantly the anomalous Hall coefficient in
several Heusler compounds.\cite{KDT13}

We show in \FG{DOSCo2TiZ} the spin-resolved 
partial and total density of states (DOS) curves for Co$_2$TiSi and
Co$_2$TiGe calculated at their respective equilibrium lattice
constants. This figure clearly demonstrates the half-metallic
character of both systems, with the Fermi energy lying close 
to the upper edge of the band gap appearing in the minority 
spin channel. The width of this gap, estimated from the calculated 
band structure (see below) is $0.834$~eV for Co$_2$TiSi and
$0.636$~eV for Co$_2$TiGe.
Our results shown in \FG{DOSCo2TiZ}, 
again in very good agreement with those obtained 
by other authors,\cite{SSK10,BFB+10} 
further evidence that the $3d$ states of Ti (red/light gray lines)
are almost completely empty and that the spin magnetic moment of
$2~\mu_{\rm B}$ found in these systems stems exclusively from the
two Co atoms (dashed blue/dark gray lines). We have performed 
a series of calculations with a stepwise increase of the Fermi
smearing to mimic the increase of the {\em electronic} temperature. It was
found that the total magnetization calculated this way
remains unchanged up to a value of $600$~K, much higher than 
the measured Curie temperature $T_C$ of these compounds, 
which lies around $380$~K.\cite{BFB+10} 
This is an indication of the fact that the
transition to the paramagnetic state is caused by a loss of 
magnetic order while the local magnetic 
moments at the Co atoms
may persist even above $T_C$.
We note, however, that the 
measurements of Barth \ea\cite{BFB+10} 
show a sharp drop in the magnetization only
at temperatures above $300$~K. It is therefore expected that
spin fluctuations as a potential source of electron scattering
are not very pronounced below room temperature.

%FFFFFFFFFFFFFFFFFFFFFFFFFFFFFFFFFFFFFFFFFFFFFFFFFFFFFFFFFFFFFFFFFFFFFFFFFFFFF
%FFFFFFFFFFFFFFFFFFFFFFFFFFFFFFFFFFFFFFFFFFFFFFFFFFFFFFFFFFFFFFFFFFFFFFFFFFFFF
%FFFFFFF
\begin{figure}
  \centering
 \includegraphics[width=8.2cm]{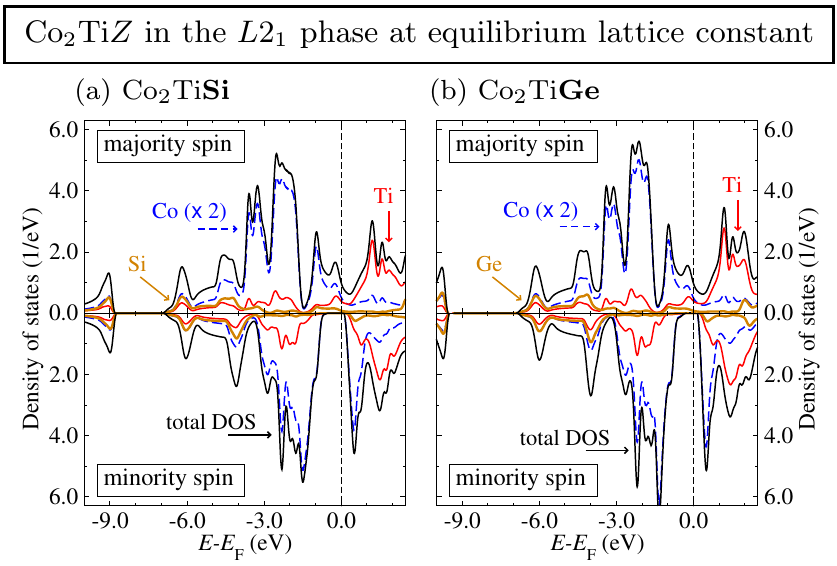}
 \caption{(Color online) Spin-polarized total and partial density of states
   of (a) Co$_2$TiSi and (b) Co$_2$TiGe as calculated for
   their cubic $L2_1$ phase at the corresponding 
   equilibrium lattice constants given in 
   Table~\protect\ref{TabGSabEQ}.}
 \label{DOSCo2TiZ}
\end{figure}
%FFFFFFF
%FFFFFFFFFFFFFFFFFFFFFFFFFFFFFFFFFFFFFFFFFFFFFFFFFFFFFFFFFFFFFFFFFFFFFFFFFFFFF
%FFFFFFFFFFFFFFFFFFFFFFFFFFFFFFFFFFFFFFFFFFFFFFFFFFFFFFFFFFFFFFFFFFFFFFFFFFFFF

\subsection{Free standing epitaxial Heusler alloys on Al(001)}
           \label{SubSecEpiHeu}

We describe the epitaxial Heusler alloys by tetragonal structures 
with a lattice constant $a_0 = 4.027$~\AA, corresponding to
the GGA-PBE equilibrium value for Al obtained from the 
used pseudopotential. The corresponding 
$L2_1$ lattice constant is $a_{\rm epi}(L2_1)=a_0\sqrt{2}=5.695$~\AA,
which must be compared with the equilibrium lattice constants
of the cubic $L2_1$ structure given in
Table~\ref{TabGSabEQ}.
Since $a_{\rm epi}(L2_1)$ is smaller than $a(L2_1)$ for both systems
($1.07$~\% and $2.70$~\% mismatch, respectively),
the epitaxial matching is expected to produce a compressive strain 
which leads to a tetragonal distortion with a 
$c/a(L2_1)$ ratio larger than one.

The results of our calculations are summarized in
\FG{EandMEpiHeu} where we show the total energy (left side axis) and total
magnetization (right side axis) for (a) Co$_2$TiSi, and (b)
Co$_2$TiGe. The calculated equilibrium $c/a(L2_1)$ ratios are
$1.022$ for Co$_2$TiSi and 
$1.063$ for Co$_2$TiGe. These values are, as expected, larger
than one, and consistent with the larger mismatch
and the smaller bulk modulus $B$ of Co$_2$TiGe.
We also note that the individual
atomic displacements during the internal relaxation 
did not exceed $10^{-3}$\AA\ with respect to the 
symmetric positions, preserving an
equally spaced $c/4$ vertical stacking.
 
%FFFFFFFFFFFFFFFFFFFFFFFFFFFFFFFFFFFFFFFFFFFFFFFFFFFFFFFFFFFFFFFFFFFFFFFFFFFFF
%FFFFFFFFFFFFFFFFFFFFFFFFFFFFFFFFFFFFFFFFFFFFFFFFFFFFFFFFFFFFFFFFFFFFFFFFFFFFF
%FFFFFFF
\begin{figure}[t]
  \centering
 \includegraphics[width=8.2cm]{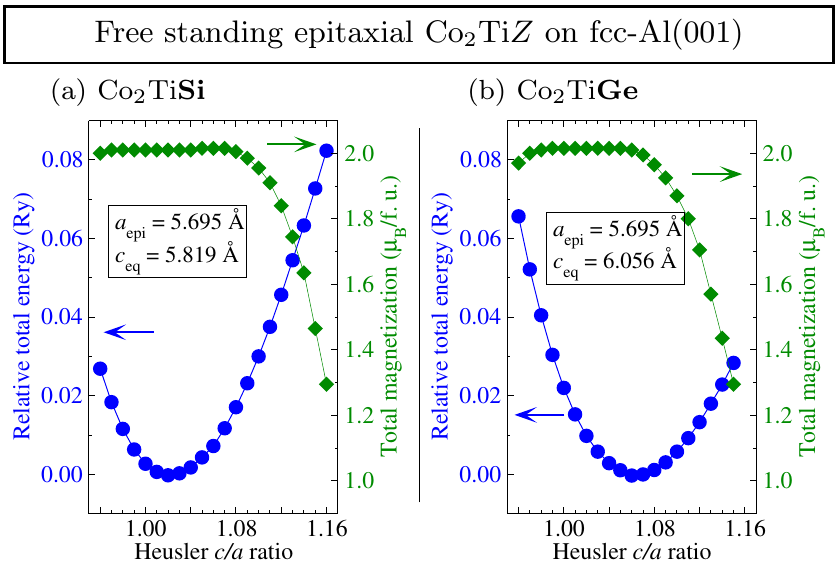}
 \caption{(Color online) Relative total energy with respect to its
   minimum (circles, left side axis)
   and total magnetization (diamonds, right side axis) for the tetragonally 
   distorted (a)~Co$_2$TiSi and (b)~Co$_2$TiGe with a fixed in-plane 
   lattice constant $a_0=4.027$~\AA. The inset gives the
   equilibrium parameters in the appropriately distorted $L2_1$
   structure.}
 \label{EandMEpiHeu}
\end{figure}
%FFFFFFF
%FFFFFFFFFFFFFFFFFFFFFFFFFFFFFFFFFFFFFFFFFFFFFFFFFFFFFFFFFFFFFFFFFFFFFFFFFFFFF
%FFFFFFFFFFFFFFFFFFFFFFFFFFFFFFFFFFFFFFFFFFFFFFFFFFFFFFFFFFFFFFFFFFFFFFFFFFFFF

From \FG{EandMEpiHeu} it is also evident that both materials 
remain half-metallic (the total magnetization is 
$\simeq 1.00$~$\mu_{\rm B}$/Co atom) 
over a broad range of $c/a(L2_1)$ ratios. Moreover, the drop in the 
total magnetization occurs {\em above} the
$c/a(L2_1)$ equilibrium ratio. Note that the local spin magnetic
moments on the Ti and $Z$ sites are negligible small.
The preservation of half-metallicity 
at relatively large tetragonal distortions will prove to be an 
important result in view of the transport properties of the Al/Heusler/Al
trilayers. 

Here we have to emphasize that, since we considered a
single in-plane lattice constant, the attained minimum 
of the total energy should not be understood as a proof 
for the existence of a stable tetragonal structure for the 
considered Heusler alloys. In fact, Meinert \ea\cite{MSR10} performed
total energy calculations for the isoelectronic system 
Co$_2$TiSn over a series of in-plane lattice constants. 
In the absence of an epitaxial constraint, no tetragonal structure 
was found to have smaller total energy than the cubic $L2_1$ structure. 

The effect of the biaxial strain on the band structure of the
Co$_2$Ti$Z$ is
% rather
significant, and the way in which
the individual bands are affected is non-trivial. 
We illustrate this in \FG{EvsKinEPI}
for the Co$_2$TiGe Heusler alloy. Qualitative similar
features were found also for Co$_2$TiSi. In this case, however, 
the biaxial strain is smaller and the corresponding effects
on the band structure are weaker.

%FFFFFFFFFFFFFFFFFFFFFFFFFFFFFFFFFFFFFFFFFFFFFFFFFFFFFFFFFFFFFFFFFFFFFFFFFFFFF
%FFFFFFFFFFFFFFFFFFFFFFFFFFFFFFFFFFFFFFFFFFFFFFFFFFFFFFFFFFFFFFFFFFFFFFFFFFFFF
%FFFFFFF
\begin{figure*}
  \centering
 \includegraphics[height=6.5cm]{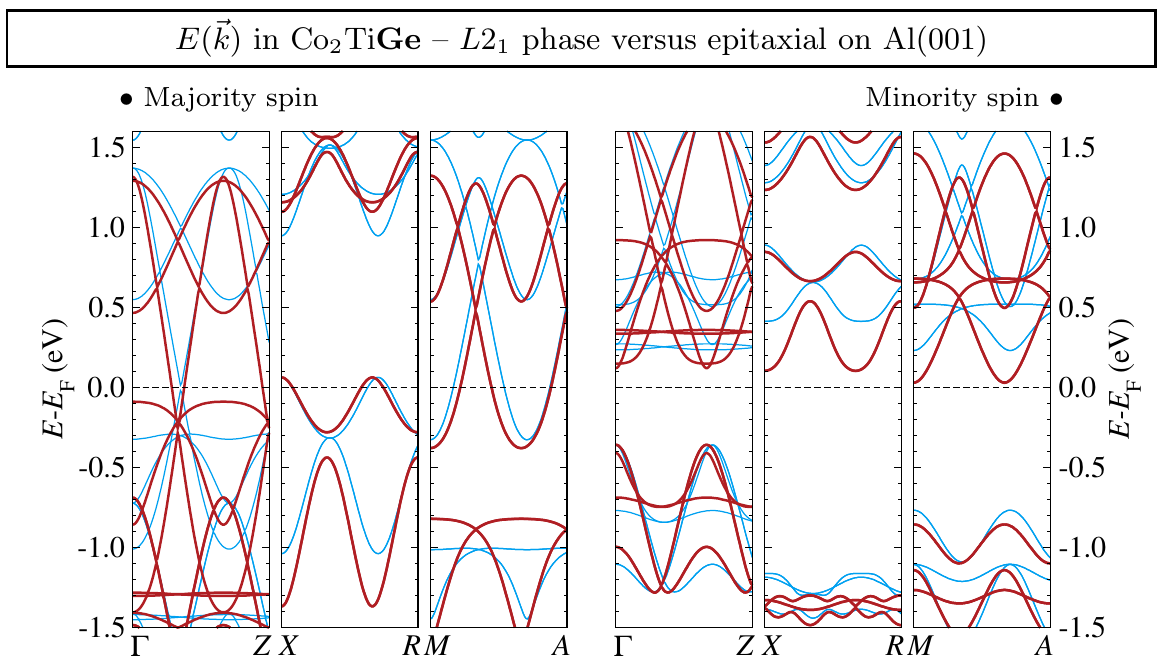}
 \caption{(Color online) Comparison of the spin-resolved 
          band structure [majority (minority) spin in the left (right)
          panel] calculated for
          the Heusler alloy Co$_2$TiGe in the equilibrium $L2_1$~phase
          [thin light blue (grey) lines] versus that of the 
          tetragonally distorted Al(001)-matched structure [thick dark
          red (grey) lines]. Both calculations were done using a
          tetragonal unit cell to which the notation of the high-symmetry
          points in the BZ refers. Note that for each
          selected direction only the $k_z$
          component of the wavevector varies, while
          $k_x$ and $k_y$ are fixed.}
      \label{EvsKinEPI}
\end{figure*}
%FFFFFFF
%FFFFFFFFFFFFFFFFFFFFFFFFFFFFFFFFFFFFFFFFFFFFFFFFFFFFFFFFFFFFFFFFFFFFFFFFFFFFF
%FFFFFFFFFFFFFFFFFFFFFFFFFFFFFFFFFFFFFFFFFFFFFFFFFFFFFFFFFFFFFFFFFFFFFFFFFFFFF

Figure~\ref{EvsKinEPI} shows the spin-resolved 
dispersion relations $E(\vec k)$, with the two panels,
left and right, displaying respectively the 
majority and minority spin bands. For each spin, the
bands stemming from the two structures, $L2_1$ 
and epitaxial tetragonal, are put together in the same frame. 
In both cases the band structure calculations employed 
a tetragonal unit cell of similar construction as above,
with lattice constants adopted to recover the appropriate
geometry. Adopting a Bravais lattice of the same
symmetry has the advantage of dealing with 
identical {\em folding} of the bands, which allows one to directly
isolate those changes that are solely due to the biaxial strain. 
Further comparison with 
band structure literature data\cite{BFB+10,SSK10} is 
possible, but demanding.\cite{FoldingNote}

The $E(\vec k)$ relations are represented in \FG{EvsKinEPI}
for several directions in the BZ for
which only the $k_z$ component of $\vec k$
varies, while $k_x$ and $k_y$ are fixed: $\Gamma$-$Z$, 
$X$-$R$, and $M$-$A$. The notation used here 
corresponds to the tetragonal BZ.\cite{SC10}
In addition, the band structure is shown over a small 
energy interval ($3$~eV wide) around the Fermi
energy $E_{\rm F}$. 
The value of $E_{\rm F}$, different in the two phases, is taken here as 
reference. This choice, while still enabling the strain effect
analysis, will also prove useful when
discussing the electronic transmission probability 
in the Al/Heusler/Al trilayers.

For the cubic $L2_1$ structure [thin light blue (grey) lines
in \FG{EvsKinEPI}] only one band crosses the Fermi energy in the
majority spin channel (left panel). 
This band is accompanied by its 
fcc-folded pair, particularly evident along the 
$\Gamma-Z$ and $M-A$ directions. 
Most of the majority spin bands in the chosen energy interval
are highly dispersive. Several localized Co $d$~bands can 
be observed, e.g., the flat bands at $-0.3$~eV and $-1.5$~eV 
along $\Gamma-Z$ and around $-1.0$~eV along $M-A$. 
The bands of the epitaxial tetragonal Co$_2$TiGe 
[thick dark red (grey) lines] are shifted in energy against
that of the cubic system but each by a different amount.
While all the $d$~bands appear to move towards
higher energies under biaxial strain, the shift of the
$s$ and $p$ bands depends both on energy and on $\vec k$.
For example, the spin-up band crossing $E_{\rm F}$ along 
$M-A$ is hardly affected by the tetragonal distortion, 
whereas the one right below it is found much lower in 
energy than its $L2_1$ counterpart. A close inspection 
of the other panels, e.g., below $E_{\rm F}$ along $\Gamma-Z$, 
reveals similar characteristics, indicating a non-rigid
band structure shift under biaxial strain. 

Similar changes of the band structure with
the tetragonal distortion are also observed in the 
minority spin channel (right panel). Here the most striking
feature is how the band gap in this spin channel 
-- the landmark of half-metallicity -- strongly diminishes 
under epitaxial strain. It can be seen, in fact, that the 
'minority conduction band' in epitaxial Co$_2$TiGe nearly 
touches the Fermi energy along $M-A$ direction, while the 
'valence band' is practically pinned relative to $E_{\rm F}$. 
This provides the explanation for the drop in the 
total magnetization evidenced in \FG{EandMEpiHeu}: Since with 
increasing height of the tetragonal cell the minority spin band 
gap gets narrower and it plunges below the Fermi energy, 
the system becomes metallic. Note that the tetragonal distortion
of epitaxial Co$_2$TiSi at equilibrium  is less pronounced and,
as a consequence, the minority spin band gap reduction is
smaller in this case. 

We will investigate in the next section the Al/Heusler interface
and will show that, within few monolayers away from it, 
the Heusler alloys stabilize into a tetragonal structure of 
identical geometry as determined here for the free standing system. 
One can therefore regard the Al/Heusler/Al trilayer systems as 
comprising a tetragonally distorted Heusler alloy spacer. 
The band structure features discussed here have an important 
influence on the transport properties of the whole junction; 
for example, the minority-spin band gap reduction is equivalent to a 
lowering of the potential barrier for spin-down electrons.

%SSSSSSSSSSSSSSSSSSSSSSSSSSSSSSSSSSSSSSSSSSSSSSSSSSSSSSSSSSSSSSSSSSSSSSSSSSSSSSS
%SSSSS
%SSSSS
\section{Stability and Properties of the 
Al/Co\bm{_2}Ti\bm{Z} Interface}\label{SecIface}

The matching of Al(001) and Co$_2$Ti$Z$ Heusler alloys in 
a heterostructure appears justified by the morphology
considerations made above. Our arguments were based so far
on the atomic arrangement and the small lattice mismatch between the
two systems. We will show in this section by means
of {\em ab initio} thermodynamics that the
Al/Co$_2$Ti$Z$ ($Z=$ Si, Ge) interfaces, both with Co-Co and
Ti-$Z$ terminations, have a negative formation energy
and are thus stable against Al and Co$_2$Ti$Z$ separation.
The stabilization of Ti-$Z$-terminated interfaces, however,
requires non-equilibrium growth conditions because of
the competing formation of Ti$Z$ compounds.
In addition, we find that within a short distance 
away from the interface the Heusler systems take on 
the epitaxial geometry discussed in the preceding section, 
with a preserved half-metallicity reflected
in the electronic structure. Such a fast transition to the 
half-metallic state appears to be an ubiquitous characteristic 
of the Heusler-based interfaces, whether they
contain other Heusler alloys\cite{CGC+11}, semiconductors like
Si,\cite{WKS05} or insulators like MgO.\cite{HSK09}

%sssssssssssssssssssssssssssssssssssssssssssssssssssssssssssssssssssssssssssssss
%sssss
%sssss
\subsection{Stability of the interface}

We describe the interfaces in our calculations using the 
tetragonal supercells depicted in \FG{FIGStrucs}(b) and (c),
assuming that the Al fcc substrate lattice can be continued
either by Ti and $Z$ atoms ($Z$= Si/Ge),  
or by two Co atoms.  
Based on total energy calculations including structural 
relaxation, we calculate the interface energy as a function 
of the reduced chemical potential of Co,
$\tilde\mu_{\text{Co}} = \mu_{\text{Co}} - E_{\text{Co}}$,
according to Section~\ref{SubsecIFACE}.

The results of our calculations are displayed in 
\FG{AlHeuEiface}(a) and (b).
Each panel shows two lines corresponding to the
Ti-$Z$ and Co-Co termination of the respective interface,
Al/Co$_2$TiSi [panel (a)] and Al/Co$_2$TiGe [panel (b)].
As can be seen, the two systems exhibit a common behavior: 
the formation energy of the Co-Co terminated interface
is smaller for Co-rich conditions 
($\tilde\mu$ approaching zero), whereas in Co-poor
conditions it is the Ti-$Z$ terminated interface
having a smaller energy. In both cases the two lines
intersect {\em below} zero. This
means that, in principle, either of the two terminations 
can be stabilized against Al and Co$_2$Ti$Z$ separation,
depending on $\tilde\mu_{\rm Co}$, and thus 
on the growth conditions.

The actual stabilization of a particular interface, on the other
hand, will be influenced by the competition with other stable products
that may appear during the preparation process. As discussed in
Section~\ref{SubsecIFACE}, the upper bound of 
$\tilde\mu_{\text{Co}}$ is $0.0$~eV, corresponding to the formation of
metallic Co. Lower bounds for $\tilde\mu_{\text{Co}}$ can be obtained
from \EQ{mucolow} or analogous expressions, assuming 
the various elements Co, Ti, and $Z$ in thermodynamic equilibrium
with competing compounds such as CoTi, Ti$Z$, and Co$Z$. The 
evaluation of these lower bounds requires total energy DFT 
determinations for the different components in their ground state. 
For $Z~=$~Si such calculations were already done by the
authors\cite{GK13} using the current pseudopotentials and
exchange-correlation functional. A similar
procedure was applied to CoTi (CsCl structure), 
CoGe ($B20$ structure), and TiGe ($B27$ structure).
Regarding the latter, 
we have searched over several possible structures and found 
the $B27$ to have the smallest total energy and being stable
against Ti and Ge segregation, in analogy to the TiSi compound.\cite{GK13} 

%FFFFFFFFFFFFFFFFFFFFFFFFFFFFFFFFFFFFFFFFFFFFFFFFFFFFFFFFFFFFFFFFFFFFFFFFFFFFF
%FFFFFFFFFFFFFFFFFFFFFFFFFFFFFFFFFFFFFFFFFFFFFFFFFFFFFFFFFFFFFFFFFFFFFFFFFFFFF
%FFFFFFF
\begin{figure}
  \centering
  \includegraphics[width=8.2cm]{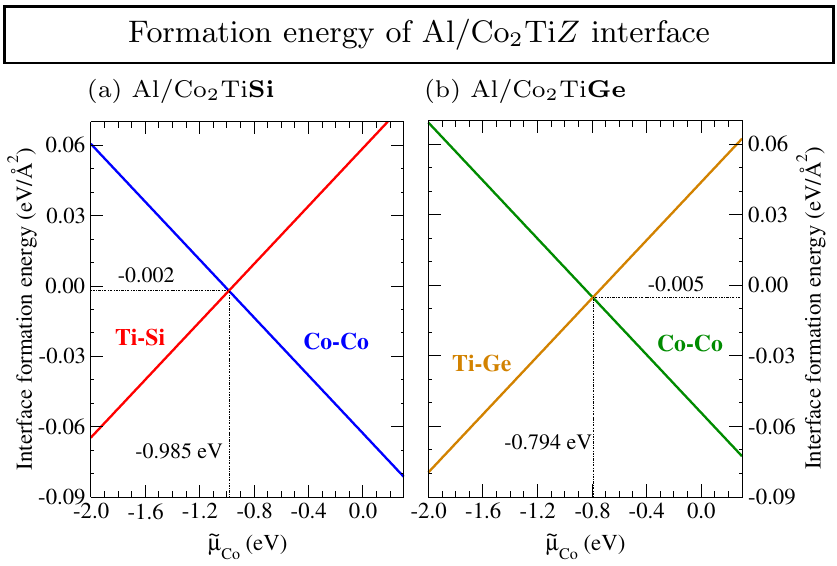}
    \caption{(Color online) Interface formation energy for the epitaxial
    (a) Al/Co$_2$TiSi 
    and (b) Al/Co$_2$TiGe interfaces, calculated 
   as a function of the reduced chemical potential of Co,
   $\tilde\mu_{\text{Co}} = \mu_{\text{Co}} - E_{\text{Co}}$.
   In each panel, the two lines correspond to the different
   interface terminations, Co-Co or Ti-$Z$. The coordinates of
   the crossing points are also provided.}
  \label{AlHeuEiface}
\end{figure}
%FFFFFFF
%FFFFFFFFFFFFFFFFFFFFFFFFFFFFFFFFFFFFFFFFFFFFFFFFFFFFFFFFFFFFFFFFFFFFFFFFFFFFF
%FFFFFFFFFFFFFFFFFFFFFFFFFFFFFFFFFFFFFFFFFFFFFFFFFFFFFFFFFFFFFFFFFFFFFFFFFFFFF

Taking into account the formation of CoTi, Ti$Z$, and Co$Z$ leads to
the following lower bounds of $\tilde\mu_{\text{Co}}$ for 
the Al/Co$_2$TiSi interface:
\begin{equation}
  \label{AlCo2TiSibounds}
  \tilde\mu_{\text{Co}} \geq \left\{
    \begin{array}{ll}
      -\tilde\mu_{\text{Si}} - 1.83\mbox{ eV} & \mbox{for CoTi}\\
      -\tilde\mu_{\text{Ti}} - 1.45\mbox{ eV} & \mbox{for CoSi}\\
      -0.54\mbox{ eV} & \mbox{for TiSi} 
    \end{array}\right. \enspace,
\end{equation}
whereas for the Al/Co$_2$TiGe interface we obtain
\begin{equation}
  \label{AlCo2TiGebounds}
  \tilde\mu_{\text{Co}} \geq \left\{
    \begin{array}{ll}
      -\tilde\mu_{\text{Ge}} - 1.15\mbox{ eV} & \mbox{for CoTi}\\
      -\tilde\mu_{\text{Ti}} - 1.60\mbox{ eV} & \mbox{for CoGe}\\
      -0.52\mbox{ eV} & \mbox{for TiGe} 
    \end{array}\right. \enspace.
\end{equation}
Here, analogous to Co metal, 
$\tilde\mu_{\text{Ti}}$ and $\tilde\mu_Z$ have $0.0$~eV
as upper bounds, set by the formation of Ti, Si, and Ge 
bulk material.

Comparing these values with the Ti-$Z$/Co-Co crossing points
in \FG{AlHeuEiface}, it becomes apparent that Ti$Z$ 
spontaneous formation can occur in the range
of $\tilde\mu_{\text Co}$ where Ti-$Z$-terminated interfaces
are stabilized. Non-equilibrium growth conditions are therefore
necessary if such interfaces are desired.

%sssssssssssssssssssssssssssssssssssssssssssssssssssssssssssssssssssssssssssssss
%sssss
%sssss
\subsection{Atomic displacements near the interface}

The atomic configurations obtained during the total energy 
minimization procedure of the interface systems
are depicted schematically in \FG{AlHeuIfaceDisplace}. 
While the two types of Heusler terminations
(Co-Co or Ti-$Z$ plane) are expected to be different, 
both systems considered here, Al/Co$_2$TiSi and Al/Co$_2$TiGe, 
share some qualitative features that will be
briefly discussed in the following.
Quantitative differences in the
atomic displacements arise from the different 
equilibrium bond lengths in the two
Heusler materials. 

%FFFFFFFFFFFFFFFFFFFFFFFFFFFFFFFFFFFFFFFFFFFFFFFFFFFFFFFFFFFFFFFFFFFFFFFFFFFFF
%FFFFFFFFFFFFFFFFFFFFFFFFFFFFFFFFFFFFFFFFFFFFFFFFFFFFFFFFFFFFFFFFFFFFFFFFFFFFF
%FFFFFFF
\begin{figure}
  \centering
  \includegraphics[width=8.2cm]{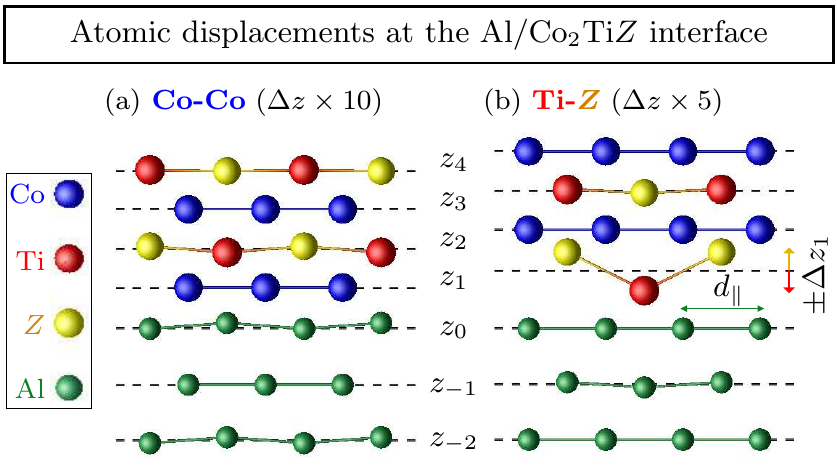}
    \caption{(Color online) 
             (110)-projected, schematic representations of the inter-atomic
             distances after accounting for relaxation at
             the epitaxial Al/Co$_2$Ti$Z$ interface with
             (a) Co-Co and (b) Ti-$Z$ terminations, defining
             the $z_i$ and $\Delta z_i$ referred to in text.
             While the inter-planar and lateral
             separations are represented on a realistic scale,
             the atomic vertical displacements are enhanced 
             by a factor 
             $s=10$ [panel (a)] or $s=5$ [panel (b)],
             labeled as $\Delta z\times s$. 
             One notes a relaxation-induced corrugation
             of the Ti-$Z$ planes, while the Co-Co planes remain flat.
             The in-plane inter-atomic
             distance has a fixed, system-independent value 
             of $d_\parallel=a_0\sqrt{2}/2=2.847$~\AA,
             with $a_0$ being the equilibrium lattice constant of fcc-Al.
             }
  \label{AlHeuIfaceDisplace}
\end{figure}
%FFFFFFF
%FFFFFFFFFFFFFFFFFFFFFFFFFFFFFFFFFFFFFFFFFFFFFFFFFFFFFFFFFFFFFFFFFFFFFFFFFFFFF
%FFFFFFFFFFFFFFFFFFFFFFFFFFFFFFFFFFFFFFFFFFFFFFFFFFFFFFFFFFFFFFFFFFFFFFFFFFFFF

Figure~\ref{AlHeuIfaceDisplace} 
shows the Al/Heusler interface in the immediate vicinity
of the contact surface, as seen in a (110)-projection 
relative to the tetragonal supercell [equivalent to the
(100)-direction of the $L2_1$ structure]. 
Since each atomic plane contains two inequivalent 
sites $A$ and $B$, we define the $z_I$ coordinate of plane $I$ 
as the average $(z_A+z_B)/2$. 
Then, the $z$-coordinates of sites $A$ and $B$ can be given 
relative to $z_I$ through the displacement $\Delta z_I=|z_A-z_B|/2$
as $z_{A,B}=z_I\pm\Delta z_I$. The plane coordinates 
can also be used to define the inter-planar
distance between two successive atomic planes
as $d_I=|z_I-z_{I-1}|$. Relevant values in the 
vicinity of the interface are listed
in Table~\ref{TabDisplace}. 
Note that the reference $z_0=0.0$ was taken 
for the topmost Al-Al plane.

%TTTTTTTTTTTTTTTTTTTTTTTTTTTTTTTTTTTTTTTTTTTTTTTTTTTTTTTTTTTTTTTTTTTTTTTTTTTTT
%TTTTTTTTTTTTTTTTTTTTTTTTTTTTTTTTTTTTTTTTTTTTTTTTTTTTTTTTTTTTTTTTTTTTTTTTTTTTT
%TTTTTTTTTTTTT
\begin{table}
  {\tabcolsep0.5ex
  \begin{tabular}{lrrrrr}\hline\hline
    System: & \multicolumn{2}{c}{Al/Co$_2$TiSi} & \rule{2ex}{0pt} &
              \multicolumn{2}{c}{Al/Co$_2$TiGe} \\
    Termination: & \multicolumn{1}{c}{Co-Co} &
            \multicolumn{1}{c}{Ti-Si} & & 
            \multicolumn{1}{c}{Co-Co} &
            \multicolumn{1}{c}{Ti-Ge} \\\hline\hline
%%%%%%%%%%%%%%%
%%%%%%%%%%%%%%%%%%%%
    $d_z^{\,\rm epi}$ & \multicolumn{2}{c}{$1.455$} & & 
                    \multicolumn{2}{c}{$1.514$} \\\hline
%%%%%%%%%%%%%%%
    $d_4$     & 1.431 & 1.468 & & 1.493 & 1.524 \\
%%%%%%%%%%%%%%%
    $d_3$     & 1.478 & 1.419 & & 1.527 & 1.470 \\
%%%%%%%%%%%%%%%
    $d_2$     & 1.410 & 1.536 & & 1.498 & 1.609 \\
%%%%%%%%%%%%%%%
    $d_1$     & 1.517 & 2.120 & & 1.490 & 2.192 \\
%%%%%%%%%%%%%%%
    $d_{-1}$   & 2.071 & 2.068 & & 2.077 & 2.056 \\
%%%%%%%%%%%%%%%
    $d_{-2}$ & 2.041 & 2.026 & & 2.047 & 2.022 \\\hline
%%%%%%%%%%%%%%%
    $a_0/2$  & \multicolumn{4}{c}{2.013} & \\\hline\hline
%%%%%%%%%%%%%%%%%%%%%%%%%%%%%%
  \end{tabular}}
  \caption{Inter-planar separations %$d_I=|z_I-z_{I-1}|$ 
    (in \AA) for
    the relaxed epitaxial interfaces
    Al/Co$_2$TiSi and Al/Co$_2$TiGe and for 
    the different terminations. 
    The equilibrium inter-planar distances 
    for the epitaxial
    Heusler alloys ($d_z^{\,\rm epi}=c_{\rm epi}/4$)
    and for fcc-Al ($a_0/2$)
    are also given for comparison.}
    \label{TabDisplace}
\end{table}
%TTTTTTTTTTTTT
%TTTTTTTTTTTTTTTTTTTTTTTTTTTTTTTTTTTTTTTTTTTTTTTTTTTTTTTTTTTTTTTTTTTTTTTTTTTTT
%TTTTTTTTTTTTTTTTTTTTTTTTTTTTTTTTTTTTTTTTTTTTTTTTTTTTTTTTTTTTTTTTTTTTTTTTTTTTT

The most important features derived from our
calculations can be summarized as follows: (i) while
the Co-Co chains remain coplanar, the Ti-$Z$ ones 
get corrugated. Two successive Co-Co planes can be seen
as building a body-centered tetragonal (bct) unit cell
in the center of which alternating Ti and $Z$ atoms are
placed. In the vicinity of the interface the 
Ti are being pulled much stronger towards the Al 
substrate than the $Z$ atoms, giving rise to
a zig-zag arrangement of the Ti-$Z$ chains.
The maximum Ti-$Z$ vertical displacement is obtained for
the first Ti-$Z$ plane, $\Delta z_1=0.14$ or
$0.13$~\AA, for $Z~=$~Ge or Si, respectively.
(ii) the Al-Heusler separation strongly 
depends on the interface termination, but shows
only a weak variation with the type of Heusler at a 
given termination. This quantity corresponds to
$d_1$ in Table~\ref{TabDisplace} and 
shows that the Co-Co planes come much closer
to the substrate than the Ti-$Z$ ones.
(iii) at $3-4$~monolayers (MLs) 
away from the interface, slightly varying with
the termination, the $\Delta z$'s  are zero and 
the inter-planar spacings equal those of the free standing 
epitaxial Heusler alloy. Here we understand by one ML two
successive atomic planes.
Table~\ref{TabDisplace} gives the inter-planar separations up
to $2$~MLs from the interface. These are compared with the corresponding
values in bulk Al ($a_0/2$, substrate) and free standing epitaxial 
Co$_2$Ti$Z$ ($d_z^{\,\rm epi}=c_{\rm epi}/4$). It can be seen that
already within the second ML these values are quite close, 
leading to the conclusions that (a) deeper Al layers are in the
almost perfect
fcc structure, and (b) the epitaxial Co$_2$Ti$Z$ material 
is distorted only in the immediate proximity of the interface.

%sssssssssssssssssssssssssssssssssssssssssssssssssssssssssssssssssssssssssssssss
%sssss
%sssss
\subsection{Density of states and local spin magnetization}

The fast transition to a periodic geometric arrangement along the
$(001)$ direction within few MLs away from the interface is also
reflected in the electronic structure of the two investigated
systems. We illustrate this behavior with the example of 
the Al/Co$_2$TiGe interface, for which we show in 
\FG{AlHeuIfaceDOSandPOL}(a) and (b) the spin-resolved 
partial DOS for several atomic planes (each containing
two atoms) in the vicinity of 
the Co-Co- [panel (a)] and Ti-Ge-terminated [panel (b)] interfaces.
The appearance of the minority
spin band gap already $3-4$ atomic planes away from the
interface is easily recognizable in both cases. The same
applies for the Co$_2$TiSi Heusler alloy which has
an even wider band gap. 

At the bottom of panels (a) and (b) also
the partial DOS of the topmost Al plane is shown.
A direct comparison of the Al partial DOS with that of
the Co-Co plane seems to indicate a 
Heusler-substrate hybridization  
for the Co-Co termination: One notes the peak
immediately below $E_{\rm F}$ in the majority spin
channel and the one above $E_{\rm F}$ for the minority
spin DOS. A similar correspondence is absent for the
Ti-Ge-terminated interface, a direct consequence of the larger
Al/Heusler separation discussed above.

%FFFFFFFFFFFFFFFFFFFFFFFFFFFFFFFFFFFFFFFFFFFFFFFFFFFFFFFFFFFFFFFFFFFFFFFFFFFFF
%FFFFFFFFFFFFFFFFFFFFFFFFFFFFFFFFFFFFFFFFFFFFFFFFFFFFFFFFFFFFFFFFFFFFFFFFFFFFF
%FFFFFFF
\begin{figure*}
  \centering
  \includegraphics[width=\textwidth]{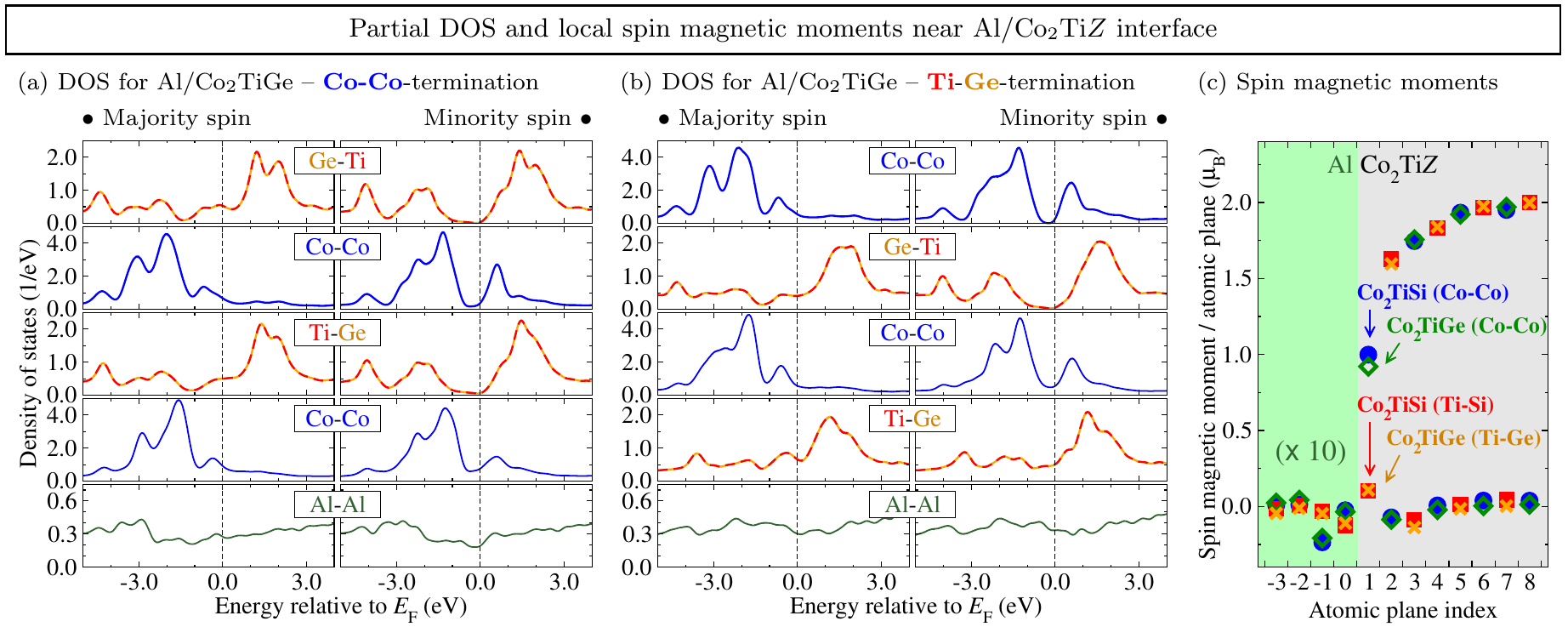}
    \caption{(Color online) Partial DOS of selected atomic planes 
      in the vicinity of the interface for the Al/Co$_2$TiGe system with
      (a) Co-Co and (b) Ti-Ge termination of the Heusler alloy.
      (c) Atomic-plane resolved spin magnetic moments
      at the Al/Co$_2$Ti$Z$
      interface for $Z~=$~Si/Ge and for both terminations. The 
      index $0$ corresponds to the topmost Al plane. Note the factor
      $10$ used to multiply the Al-related data. 
      Co atoms are found in the odd (even) positive index planes when the
      termination is Co-Co (Ti-$Z$).}
  \label{AlHeuIfaceDOSandPOL}
\end{figure*}
%FFFFFFF
%FFFFFFFFFFFFFFFFFFFFFFFFFFFFFFFFFFFFFFFFFFFFFFFFFFFFFFFFFFFFFFFFFFFFFFFFFFFFF
%FFFFFFFFFFFFFFFFFFFFFFFFFFFFFFFFFFFFFFFFFFFFFFFFFFFFFFFFFFFFFFFFFFFFFFFFFFFFF

Figure~\ref{AlHeuIfaceDOSandPOL}(c) displays the 
local spin magnetic moments, summed up over each atomic plane, 
across the Al/Co$_2$Ti$Z$ interface for both terminations. 
The labeling of the
atomic planes follows the same convention as in 
\FG{AlHeuIfaceDisplace}: The Al plane at the interface has index $0$, 
while all Co$_2$Ti$Z$ planes have positive indices. 
We further note that,
for the Co-Co (Ti-$Z$) termination, Co atoms are found in
the odd (even) planes. A \textit{very} small spin moment
($\simeq 0.001$~$\mu_{\rm B}$) is induced in the substrate, an expected
result considering the absence of (partially filled)
$d$ states in Al. Moreover, also on the Heusler
side of the interface the local magnetization is much smaller
than that of the inner layers. As expected from the partial
DOS, a fast convergence of the local spin magnetic moments with 
the plane index when moving away from the interface
is also obtained.

So far, the combined results of our investigations 
essentially  demonstrate that the effect of the 
interface on the half-metallicity of the Heusler
spacer is quite small and restricted to a very narrow 
region. When dealing with an Al/Heusler/Al trilayer 
one can in fact still regard it as a system consisting 
of a tunneling barrier for minority spin electrons.
However, its \textit{effective} thickness is smaller than 
the \textit{geometric} thickness of the Heusler film. 
In other words, the physical interface between the two 
media does not coincide with the 'electronic' interface 
separating the metallic and the half-metallic character.   

This certainly does not rule out 
other influences that may affect the size or
even the very presence of a 
minority spin band gap in these trilayers.
Among these, we mention
the effect of spin-orbit coupling on
the spin polarization\cite{MSZ+04} or
of the non-quasiparticle states 
as those found in Co$_2$MnSi.\cite{CSA+08}
While the former can be safely
ignored in the considered systems 
containing relatively light elements, the
latter might be important
for finite temperature transport properties.

%SSSSSSSSSSSSSSSSSSSSSSSSSSSSSSSSSSSSSSSSSSSSSSSSSSSSSSSSSSSSSSSSSSSSSSSSSSSSSSS
%SSSSS
%SSSSS
\section{Spincaloric effects}\label{SecSpinCal}

Ferromagnetic half-metals, such as Co$_2$TiSi and Co$_2$TiGe, offer
interesting perspectives for spincaloric applications, since they
unite features from both metals and semiconductors. On the one hand,
the minority spin channel with its energy gap may provide large
absolute values of the Seebeck coefficient, as commonly known for
semiconductors, combined with a relatively low conductivity.  On the
other hand, the metallic majority spin channel displays a larger
conductivity, albeit in conjunction with the low Seebeck coefficient
typical of a metal.  Which of the two spin contributions dominates in
the effective Seebeck coefficient is an open question that requires to
be answered for each system separately by detailed computational
studies.  For the spin Seebeck coefficient, the relative sign in
either spin channel is important, as the two spin contributions may
add up or largely cancel each other. Moreover, even in each separate
spin channel, one has to be aware of cancellation effects if
electron-like and hole-like carriers contribute in about equal
amounts.  This balance, in turn, depends sensitively on the position
of the Fermi level, as well as on the filtering effect that arises
from the variation of the transmission coefficient of the charge
carriers through the interface between the Heusler spacer layer and
the leads.

Measurements for {\em bulk} Co$_2$Ti$Z$ ($Z=$~Si, Ge, Sn) by Barth
\ea\cite{BFB+10} reported a negative effective Seebeck coefficient
$S_{\rm eff}(T)$ whose absolute value monotonously rises with
temperature, reaching values between $-31$~$\mu$V/K ($Z=$~Si) and
$-50$~$\mu$V/K ($Z=$~Sn) at and above the Curie temperature.  The
negative sign points to electrons, rather than holes, as the dominant
carriers in these bulk samples, while the large absolute value is
reminiscent of the thermoelectric behavior of semiconductors.

For the trilayers considered here, there are two major factors that
let us expect a substantially altered behavior of the Seebeck
coefficient compared to bulk: First, the Heusler spacers are subject
to different biaxial strain depending on the group IV element $Z$.
Secondly, interface scattering is anticipated to influence differently
the transmission across the junction, especially for thin Heusler
films. We have already seen that the type of interface termination
leads to different local changes in the structural, electronic, and
magnetic properties of the heterostructure.

Our explicit calculations presented below, 
based on Eqs.~(\ref{EffSeeb}) and (\ref{SpinSeeb}), will show   
that there is a significant dependence
of the effective and spin-dependent Seebeck coefficients,
$S_{\rm eff}$ and $S_{\rm spin}$,
on both the spacer material and the interface termination. 
The two materials Co$_2$Ti$Z$, with $Z=$Si, Ge, are well suited to
address these dependences: Since the electronic bands near $E_{\rm F}$
are derived from Co and Ti orbitals, the effect of the third element
$Z$ is rather indirect, as it introduces only small, predictable
changes of the strain state and the position of $E_{\rm F}$ within the
spin gap.  Therefore, selecting both the element $Z$ and the substrate
responsible for the epitaxial strain may allow one to tailor the
spincaloric transport properties of the trilayer.
Combined with the results of the previous section on the
Al/Heusler interface stability, the spincaloric properties of 
these systems can be tuned by targeted epitaxial growth combining the 
interface morphology with the biaxial strain.

We start by presenting the Seebeck coefficient results for
the Al/Co$_2$Ti$Z$/Al trilayers. We will then compare
them with calculated bulk Seebeck coefficients for 
the corresponding spacer materials. We close the section
by providing a detailed analysis of the transmission
probability in the trilayer systems, interpreting
the obtained results on the basis of  
the subtle but significant changes 
caused to this quantity by the different interface terminations.

\subsection{Effective and spin-dependent Seebeck 
                 coefficients in  Al/Co\bm{_2}Ti\bm{Z}/Al trilayers}

The calculated effective and spin-dependent Seebeck coefficients for
the two systems Al/Co$_2$TiSi/Al and Al/Co$_2$TiGe/Al with
different terminations are shown in a compact
form in \FG{AlHeuSeebeck}(a) and (b), respectively, for
temperatures up to $350$~K. Each $S(T)$ curve
is labeled accordingly using a 'spacer (termination)' notation. 

For the effective Seebeck coefficient in \FG{AlHeuSeebeck}(a) it is
easy to recognize the following sequence from large negative to
positive values: the $S_{\rm eff}(T)$ curves for the Ti-$Z$
terminations are above those for Co-Co terminations; furthermore, the
trilayers containing Co$_2$TiSi show positive (or very small) Seebeck
coefficients in the temperature range plotted in \FG{AlHeuSeebeck}(a),
whereas for Co$_2$TiGe the sign of $S_{\rm eff}(T)$ depends on the
termination.  The sequence of the curves is similar for the 
spin-dependent Seebeck coefficient, whereby for 
Al/Co$_2$TiGe/Al (Ti-Ge) $S_{\rm spin}(T)$
rises more steeply with temperature.  $S_{\rm spin}(T)$ attains
relatively large values as approaching $300$~K, being positive for
both Ti-$Z$-terminated systems and negative only for Al/Co$_2$TiGe/Al
(Co-Co), as can be seen in \FG{AlHeuSeebeck}(b). We note that the
Al/Co$_2$TiSi/Al system with Co-Co-terminated interfaces shows very
small values of both the effective 
$S_{\rm eff}(T)$ as well as spin-dependent Seebeck
coefficient $S_{\rm spin}(T)$.  The reason for 
this peculiar behavior will be made clear below.

%FFFFFFFFFFFFFFFFFFFFFFFFFFFFFFFFFFFFFFFFFFFFFFFFFFFFFFFFFFFFFFFFFFFFFFFFFFFFF
%FFFFFFFFFFFFFFFFFFFFFFFFFFFFFFFFFFFFFFFFFFFFFFFFFFFFFFFFFFFFFFFFFFFFFFFFFFFFF
%FFFFFFF
\begin{figure}
    \centering
  \includegraphics[width=8.2cm]{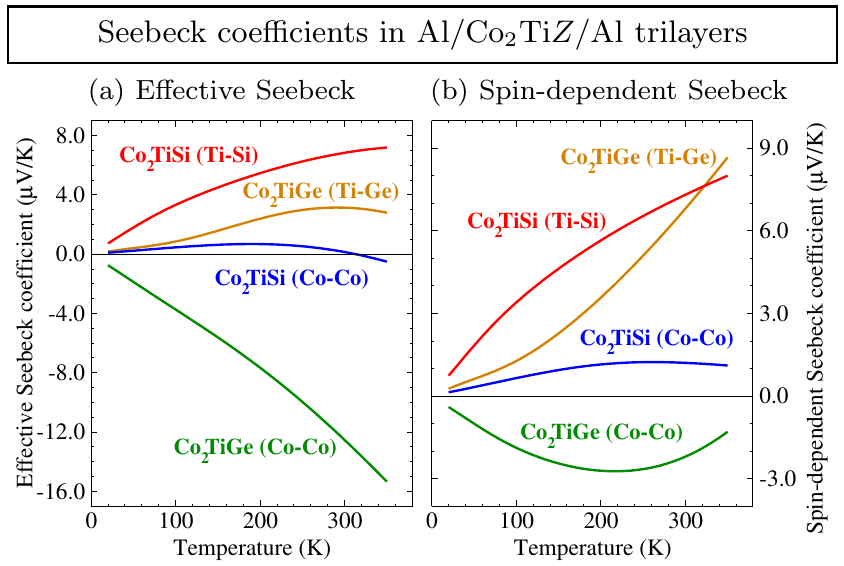}
  \caption{(Color online)
   (a) Effective and (b) spin-dependent Seebeck coefficients
   calculated according to Eqs.~(\protect\ref{EffSeeb}) and
   (\protect\ref{SpinSeeb}) for the Al/Co$_2$TiSi/Al and
   Al/Co$_2$TiGe/Al trilayer systems.
   The labeling of the curves
   designates the respective spacer material and Al-Heusler interface
   termination (Co-Co or Ti-$Z$) in parentheses.
   }
  \label{AlHeuSeebeck}
\end{figure}
%FFFFFFF
%FFFFFFFFFFFFFFFFFFFFFFFFFFFFFFFFFFFFFFFFFFFFFFFFFFFFFFFFFFFFFFFFFFFFFFFFFFFFF
%FFFFFFFFFFFFFFFFFFFFFFFFFFFFFFFFFFFFFFFFFFFFFFFFFFFFFFFFFFFFFFFFFFFFFFFFFFFFF

Equations (\ref{EffSeeb}) and (\ref{SpinSeeb}) express the effective
and spin-dependent Seebeck coefficients as a weighted sum (difference)
of the spin-resolved equivalents $S_\sigma$, defined by \EQ{Seebeck}, 
treating the two spin
channels as parallel connected resistors. Although 
the $S_\sigma$'s do not have, in a strict sense,
a physical meaning, they prove to be useful auxiliary quantities in
analyzing $S_{\rm eff}$ and $S_{\rm spin}$. 
We show the calculated $S_\uparrow(T)$ (majority spin)
and $S_\downarrow(T)$ (minority spin)
in \FG{AlHeuSebSigma}(a) and (b), where we have used the same
spacer/termination labeling convention as above.

\begin{figure}[t]
  \centering
 \includegraphics[width=8.2cm]{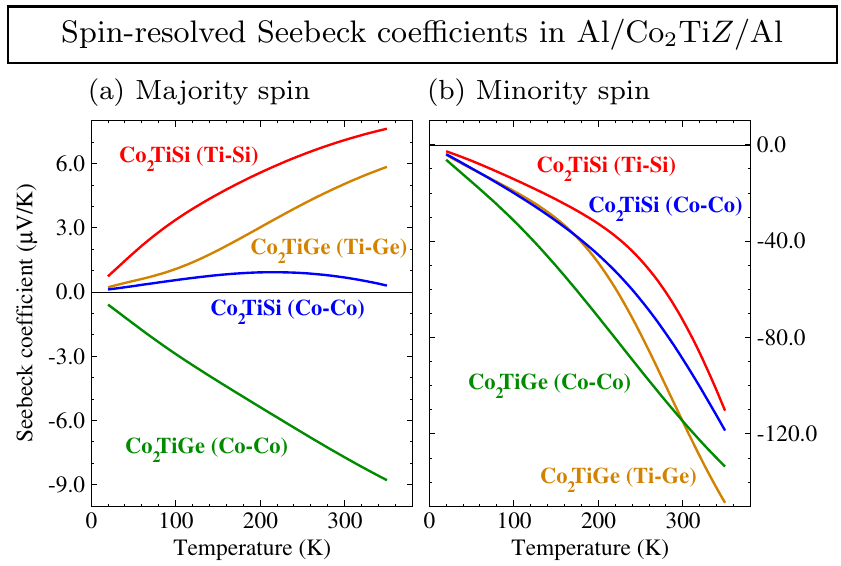}
 \caption{(Color online) Spin-resolved Seebeck coefficients,
   defined by \protect\EQ{Seebeck}, 
   for (a) majority spin and (b) minority spin channels
   calculated for the Al/Co$_2$TiSi/Al and
   Al/Co$_2$TiGe/Al trilayer systems.
   The labeling of the curves
   follows the same convention as \protect\FG{AlHeuSeebeck}.
   Note the different scales on the $y$-axis used in the two panels.}
 \label{AlHeuSebSigma}
\end{figure}

First of all, we notice that the minority spin component
$S_\downarrow$ is always negative, and much larger in absolute value
than the majority spin component $S_\uparrow$. Moreover, it exhibits a
fairly similar $T$-dependence regardless of spacer and termination.
Thus, the minority spin carriers in the Heusler spacer indeed
reproduce the thermoelectric behavior of an $n$-type semiconductor, as
could have been expected from the position of the Fermi energy in the
band gap [see \FG{EvsKinEPI}(b)].  However, since $S_\downarrow$ is
weighted with the minority spin conductance $G_\downarrow$ in
Eqs.~(\ref{EffSeeb}) and (\ref{SpinSeeb}), its contribution to both
$S_{\rm eff}$ and $S_{\rm spin}$, in spite of the very large values,
remains small. The large value of $G_\uparrow$ ensures that the
effective Seebeck coefficient is dominated by $S_\uparrow$.  This
becomes clear from a direct comparison of the (a)~panels of
\FG{AlHeuSeebeck} and \FG{AlHeuSebSigma}: The same sequence from
positive to negative values in the temperature dependence appears in
both $S_\uparrow$ and in $S_{\rm eff}$.  
Figure~\ref{AlHeuSebSigma} further shows that, except for
the Al/Co$_2$TiGe/Al (Co-Co) system, $S_\uparrow(T)$ and
$S_\downarrow(T)$ are of different sign over the whole temperature
range. Regarding the Seebeck coefficient as a potential drop, this
corresponds to electrons of different spin moving in opposite
directions across the junction, i.e., to a spin current that dominates
over the charged current.  Only for the Al/Co$_2$TiGe/Al (Co-Co)
system, {\em both} $S_\uparrow$ and $S_\downarrow$ are negative, which
largely results in a cancellation of the two contribution in 
$S_{\rm spin}$.

%FFFFFFFFFFFFFFFFFFFFFFFFFFFFFFFFFFFFFFFFFFFFFFFFFFFFFFFFFFFFFFFFFFFFFFFFFFFFF
%FFFFFFFFFFFFFFFFFFFFFFFFFFFFFFFFFFFFFFFFFFFFFFFFFFFFFFFFFFFFFFFFFFFFFFFFFFFFF
%FFFFFFF
%FFFFFFF
%FFFFFFFFFFFFFFFFFFFFFFFFFFFFFFFFFFFFFFFFFFFFFFFFFFFFFFFFFFFFFFFFFFFFFFFFFFFFF
%FFFFFFFFFFFFFFFFFFFFFFFFFFFFFFFFFFFFFFFFFFFFFFFFFFFFFFFFFFFFFFFFFFFFFFFFFFFFF

A deeper insight into the peculiarities of the Seebeck coefficients
is provided by the electronic transmission probability
${\cal T}_\sigma(E)$ defined by \EQ{TofE}, 
which lies at the core of the transport calculations. 
The way in which a specific transmission probability profile
${\cal T}(E)$ influences the sign and size of
the Seebeck coefficient can be explained in a very 
intuitive manner on the basis of \EQ{Seebeck}.\cite{CBH11}
The denominator of $S(T)$ 
in this formula (spin index omitted)
is proportional to the conductance $G(T)$, as given
by \EQ{Conduct}. Because of the $(E-E_{\rm F})$ term, the
numerator may be seen as a center of mass of 
${\cal T}(E)(\partial f_0/\partial E)$.\cite{CBH11} 
Consequently,
both sign and value of $S(T)$ will be extremely 
sensitive to small changes in the numerator's
integrand below or above $E_{\rm F}$. These changes are 
brought upon by the temperature increase which 
extends the effective non-zero width of
${\cal T}(E)(\partial f_0/\partial E)$.
This interpretation is equivalent
to Mott's formula\cite{SI86} in the limit
of $T\to 0$; that is, $S$ is proportional to 
the logarithmic derivative of the conductivity 
at $E_{\rm F}$, its sign being positive (negative)
for a negative (positive) slope of $\sigma(E)$.

The transmission probability curves calculated using
\EQ{TofE} are shown for Al/Co$_2$TiSi/Al and Al/Co$_2$TiGe/Al 
in \FG{AlHeuTrans}(a) and (b), respectively, resolved according 
to the majority (top) and minority (bottom) spin channels.

%FFFFFFFFFFFFFFFFFFFFFFFFFFFFFFFFFFFFFFFFFFFFFFFFFFFFFFFFFFFFFFFFFFFFFFFFFFFFF
%FFFFFFFFFFFFFFFFFFFFFFFFFFFFFFFFFFFFFFFFFFFFFFFFFFFFFFFFFFFFFFFFFFFFFFFFFFFFF
%FFFFFFF
\begin{figure}
  \centering
 \includegraphics[width=8.2cm]{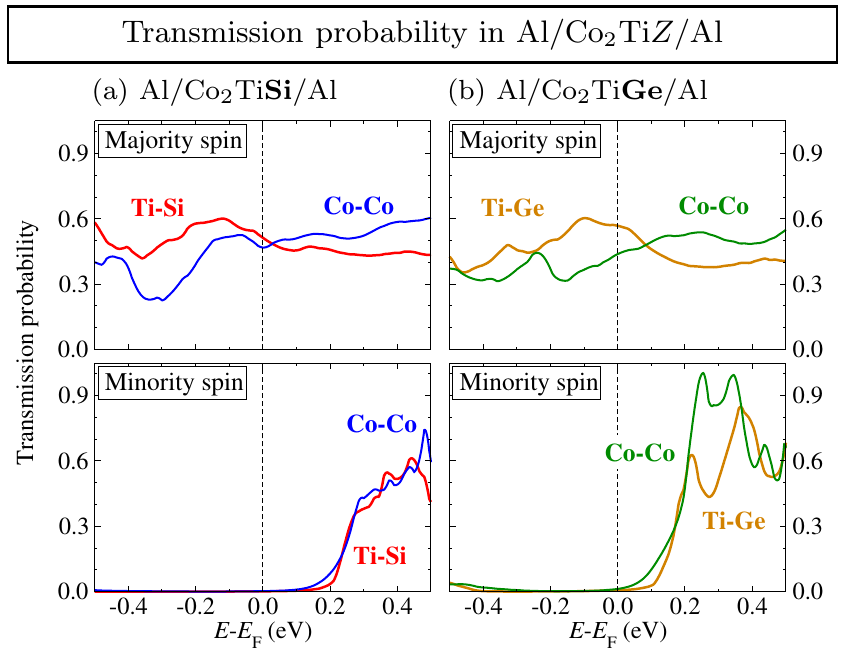}
 \caption{(Color online) Spin-resolved electronic transmission probability 
   calculated for (a)~Al/Co$_2$TiSi/Al and 
   (b)~Al/Co$_2$TiGe/Al heterostructures.
   For each system and spin component one panel contains two sets of 
   data corresponding to the different terminations, Co-Co or Ti-$Z$,
   of the Heusler-Al interface, appropriately labeled.}
 \label{AlHeuTrans}
\end{figure}
%FFFFFFF
%FFFFFFFFFFFFFFFFFFFFFFFFFFFFFFFFFFFFFFFFFFFFFFFFFFFFFFFFFFFFFFFFFFFFFFFFFFFFF
%FFFFFFFFFFFFFFFFFFFFFFFFFFFFFFFFFFFFFFFFFFFFFFFFFFFFFFFFFFFFFFFFFFFFFFFFFFFFF
In the dominant majority spin channel (upper panels) the variations of
$S_\uparrow$ between the two terminations Co-Co and Ti-$Z$ are clearly
reflected by differences in the energy-dependent transmission for both
spacers. It is, however, important to note that ${\cal T}_\uparrow(E)$
for the Ti-$Z$ terminations are quite similar in shape: both have a
broad peak right below $E_{\rm F}$ which thus lies on
a falling flank of ${\cal T}_\uparrow(E)$.  These features explain the
positive sign and the close values obtained in $S_\uparrow$, 
$S_{\rm  eff}$, and $S_{\rm spin}$ for both Ti-$Z$ trilayers.  In contrast,
the Co-Co terminations for the two spacer materials show very
different transmission profiles: 
In Al/Co$_2$TiSi/Al, ${\cal T}_\uparrow(E)$ is nearly constant 
around $E_{\rm F}$.  The Fermi energy itself lies in the middle of a 
rather symmetric dip in the transmission, which provides the reason
for the small Seebeck coefficients $S_\uparrow$, $S_{\rm eff}$, and 
$S_{\rm spin}$ obtained for Al/Co$_2$TiSi/Al (Co-Co).
For the other spacer material, Co$_2$TiGe, the maximum in transmission 
lies above $E_{\rm F}$, which is thus on a 
rising flank of ${\cal T}_\uparrow(E)$, and so all the above
coefficients are negative.

In the minority spin channel, the aforementioned gap 
(bottom panels) is correspondingly reflected in the 
transmission probability as
a broad energy interval around $E_{\rm F}$ where
${\cal T}_\downarrow(E)\to 0$. Above the Fermi energy
a sudden increase follows, with the onset higher
for Al/Co$_2$TiSi/Al than for Al/Co$_2$TiGe/Al.
This is due to the stronger tetragonal distortion in the
latter case, effectively leading to a much smaller
band gap in the strained Co$_2$TiGe spacer. Since in all
cases the transmission switches from zero, below $E_{\rm F}$,
to a finite value above it, $S_\downarrow(T)$ must necessarily be
negative. 

\subsection{Transport properties of cubic versus epitaxially 
strained bulk Co$_2$Ti\bm{Z}}

In order to understand the above results in physical terms, it is
advisable to single out possible factors contributing to the observed
trends.  One such factor, which is almost inevitable in any epitaxial
system, is epitaxial strain. It is interesting to study strain effects
in their own right, since they may provide a way to deliberately
modify the electronic structure of thin films by selecting a suitably
matched substrate.  The analysis provided below will show, however,
that, in the present case, epitaxial strain alone is not sufficient to
explain the different spincaloric properties of 
Al/Co$_2$Ti$Z$/Al trilayers.

Within our approach, we manage to distinguish the 
effects due to interface scattering from those solely due to strain 
by studying a fictitious trilayer system
having the left and right leads, as well as 
the spacer in between, consisting of identical materials. 
For each transmission channel~$\vec k_\parallel$ 
the transmission~${\cal T}(\vec k_\parallel,E)$
is a well-defined quantity, being proportional to the
$\vec k_\parallel$-projected $E(\vec k)=E(\vec k_\parallel,k_z)$
isosurface.\cite{MPD04} For example, the quantity
${\cal T}(\vec k_\parallel,E_{\rm F})$ will represent
the Fermi surface projected on the $(k_x,k_y)$-plane.

Corresponding results obtained for bulk Co$_2$TiSi and Co$_2$TiGe
are displayed in \FG{BulkTrans}. Panels~(a) of this figure show the
effective Seebeck coefficient, while panels~(b) show the
spin-resolved electronic transmission. In each case, a comparison is
made between the ideal $L2_1$ structure of the respective Heusler
alloy [thin light blue (grey) lines]
and its tetragonally distorted, Al(001) epitaxially matched
structure [thick dark red (grey) lines].

%FFFFFFFFFFFFFFFFFFFFFFFFFFFFFFFFFFFFFFFFFFFFFFFFFFFFFFFFFFFFFFFFFFFFFFFFFFFFF
%FFFFFFFFFFFFFFFFFFFFFFFFFFFFFFFFFFFFFFFFFFFFFFFFFFFFFFFFFFFFFFFFFFFFFFFFFFFFF
%FFFFFFF
\begin{figure}
  \centering
 \includegraphics[width=8.2cm]{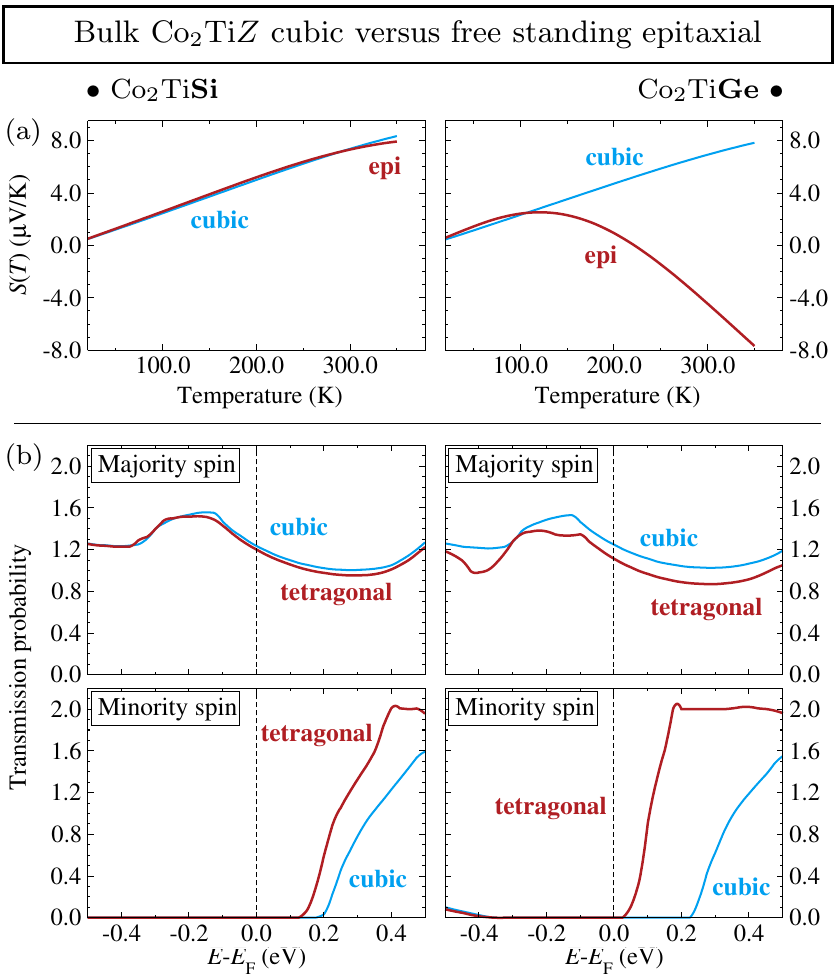}
 \caption{(Color online) (a) Effective Seebeck coefficient and (b)
   spin-resolved electronic transmission probability as calculated
   for {\em bulk} Co$_2$TiSi and Co$_2$TiGe. Each panel shows two sets
   of data, one corresponding to the $L2_1$ (cubic) Heusler structure
   (thin light blue/grey), the second to the tetragonally distorted
   structure, epitaxially matched to fcc-Al (thick dark red/grey).}
 \label{BulkTrans}
\end{figure}
%FFFFFFF
%FFFFFFFFFFFFFFFFFFFFFFFFFFFFFFFFFFFFFFFFFFFFFFFFFFFFFFFFFFFFFFFFFFFFFFFFFFFFF
%FFFFFFFFFFFFFFFFFFFFFFFFFFFFFFFFFFFFFFFFFFFFFFFFFFFFFFFFFFFFFFFFFFFFFFFFFFFFF
Our results for the effective Seebeck coefficient of the $L2_1$
Co$_2$Ti$Z$ Heusler alloys agree well with previous calculations based
on the bulk electronic structure.\cite{BFB+10,SSK10} Under epitaxial
conditions, we note that a change in $S(T)$ occurs for Co$_2$TiGe,
\FG{BulkTrans}(a), right panel, but not for for Co$_2$TiSi,
\FG{BulkTrans}(a), left panel. Only for this compound does the Fermi
energy come close enough to the conduction band such that, for finite
but not exceedingly high temperatures, the minority spin transmission
and conductance becomes finite.  As a consequence, the
semiconductor-like minority spin current $G_\downarrow S_\downarrow$
shows up in the numerator of the effective Seebeck coefficient,
\EQ{EffSeeb}, leading to the occurrence of large, negative values.

An analysis of the transmission profiles, \FG{BulkTrans}(b) for the
$L2_1$ structure shows striking similarities between Co$_2$TiSi (left)
and Co$_2$TiGe (right). This is again an indication of the fairly
identical electronic structure of the two materials around the Fermi
level, which is determined by Co- and Ti-derived electronic
bands.  However, for the tetragonally distorted (epitaxial) systems,
the transmission changes to a different extent in the two Heusler
materials because of the different biaxial strain and outward
expansion. The effect is clearly stronger for Co$_2$TiGe, as one can
see in the more pronounced shift of the minority spin transmission
towards lower energies. In fact, comparing the onsets of minority spin
transmission for the tetragonal Co$_2$TiSi and Co$_2$TiGe with the
corresponding onsets in the trilayer systems in \FG{AlHeuTrans} we
note that they match nearly perfectly. The small differences appear
because of (i) different reference Fermi energies of the leads and
(ii) the tails in ${\cal T}_\downarrow(E)$ caused by the evanescent
states at the Al/Heusler interfaces.  
By comparing ${\cal T}_\downarrow(E)$ with the band structure of 
Co$_2$TiGe in \FG{EvsKinEPI}, one can notice that the downward shift
in ${\cal T}_\downarrow(E)$ reflects the similar shift of the minority spin
bands under epitaxial distortion.  This proves once more that, even at
this small thickness of the spacer materials in Al/Co$_2$Ti$Z$/Al
trilayers, the former do act as efficient potential barriers for the
spin-down electrons by preserving their half-metallic character.

We make a final observation regarding the transmission in the majority
spin channel for both Co$_2$TiSi and Co$_2$TiGe. The two 
tetragonal bulk profiles, \FG{BulkTrans}(b), are quite 
similar to those of the Ti-$Z$-terminated trilayers, \FG{AlHeuTrans}. 
In both cases a broad peak is present
below the Fermi energy, followed by a falling transmission above
it. This has to be contrasted with the completely different pattern of
${\cal T}_\uparrow$ in the Co-Co-terminated trilayers. In
Section~\ref{SecIface} we have shown two important differences between
the two types of terminations for a given spacer material: (i) the
Co-Co-terminated Heusler alloy is much closer to the Al substrate than
the Ti-$Z$-terminated, and (ii) the partial DOS curves indicate a
Heusler/Al hybridization at the interface only in the case of a
Co-Co termination.  The clearly different Seebeck coefficients and
transmission profiles for the two terminations appear to be linked
precisely to these features.  We shall investigate in the following
the even more subtle differences in the $\vec k_\parallel$-dependence
of the transmission probability caused by the interface morphology.
Moreover, we will show how the spacer material makes a difference, even for
the same termination, due to the interplay between strain and $\vec
k_\parallel$ selectivity.

\subsection{The influence of the Al/Co\bm{_2}Ti\bm{Z} interface
                   on the transmission probability}

\begin{figure*}
  \centering
 \includegraphics[width=14.2cm]{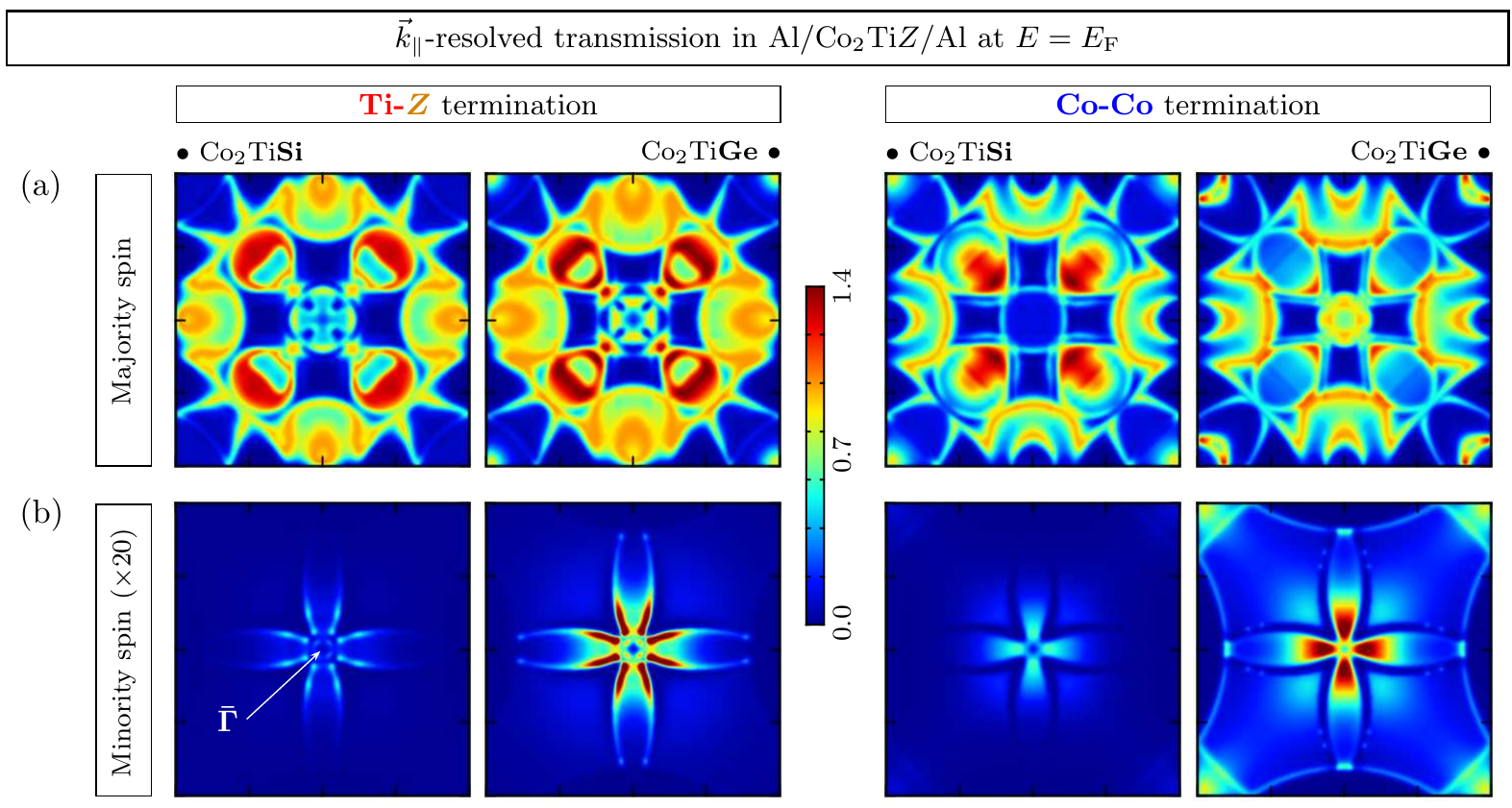}
 \caption{Contour plots of the $\vec
   k_\parallel$-resolved transmission probability in
   Al/Co$_2$Ti$Z$/Al as calculated for the energy argument 
   $E=E_{\rm F}$. Rows (a) and (b) show respectively the majority
   and minority spin channels. Left (right) columns correspond to
   the Ti-$Z$- (Co-Co-) terminated Al/Co$_2$Ti$Z$ interface.
   The BZ center is in the middle of each square, as indicated by the
   $\bar{\Gamma}$ point in the bottom-left panel. The $k_x$ ($k_y$)
   axis has horizontal (vertical) orientation.}
 \label{KapTransEF}
\end{figure*}

We investigate here the individual
contributions of the $\vec k_\parallel$ transmission channels in the
2D-BZ to the energy-resolved transmission probabilities
at selected energy arguments of \EQ{TofE}.
This proves useful in identifying which of
these channels are actually involved in transmission and to what
extent they change from one trilayer system to another.
Figure~\ref{KapTransEF} shows contour plots 
of ${\cal T}_\sigma(\vec k_\parallel,E)$ for $E=E_{\rm F}$, in the
full 2D-BZ and for all combinations of
spacer material plus interface termination studied here. 
Applying a scaling factor of $20$ to the minority spin channel transmission in
\FG{KapTransEF}(b) allows us to use a common scale for all panels.
The panels have been grouped into left and right blocks according to 
their interface termination.

%FFFFFFFFFFFFFFFFFFFFFFFFFFFFFFFFFFFFFFFFFFFFFFFFFFFFFFFFFFFFFFFFFFFFFFFFFFFFF
%FFFFFFFFFFFFFFFFFFFFFFFFFFFFFFFFFFFFFFFFFFFFFFFFFFFFFFFFFFFFFFFFFFFFFFFFFFFFF
%FFFFFFF
%FFFFFFF
%FFFFFFFFFFFFFFFFFFFFFFFFFFFFFFFFFFFFFFFFFFFFFFFFFFFFFFFFFFFFFFFFFFFFFFFFFFFFF
%FFFFFFFFFFFFFFFFFFFFFFFFFFFFFFFFFFFFFFFFFFFFFFFFFFFFFFFFFFFFFFFFFFFFFFFFFFFFF

Obvious similarities in ${\cal T}_\sigma(E)$ are
easy to recognize in the minority 
spin channel: for the Ti-$Z$-terminated interfaces (left column)
the map of Al/Co$_2$TiGe/Al (right frame) is seen to be
identical in shape with that of Al/Co$_2$TiSi/Al (left frame) and
differing only in the amplitude of the transmission at
various $\vec k_\parallel$ points, reflected by the different color. 
A similar correspondence can be seen for
the Co-Co-terminated systems in the right column of \FG{KapTransEF}(b).
As anticipated from the
${\cal T}_\downarrow(E)$ transmission profiles in \FG{AlHeuTrans}, the
${\cal T}_\downarrow(E,\vec k_\parallel)$ plots bear the typical 
characteristics of tunneling: very few transmission channels,
mostly oriented along $k_x$ and $k_y$ and placed near the
2D-BZ center (close to normal incidence), do participate in
the transmission. On the other hand, comparison of the frames 
corresponding to identical spacer, e.g., Co$_2$TiGe, but
different terminations, shows clear qualitative differences in the
transmission which are caused by the nature of the Al/Heusler contact.

The majority spin channels, \FG{KapTransEF}(a), for the two Ti-$Z$
terminations also display striking similarities: the different spacer
materials change only the amplitude in transmission, but do not
introduce or remove any individual transmission channels. This is
directly reflected in the two transmission profiles 
${\cal T}_\uparrow(E)$ (\FG{AlHeuTrans}) and the Seebeck coefficients
calculated for the Ti-$Z$ terminations (\FG{AlHeuSebSigma}) that were
found to share the same qualitative energy and temperature dependence.

On the contrary, dissimilar transmission patterns are observed for the
Co-Co-terminated Al/Co$_2$Ti$Z$/Al trilayers in the majority spin
channel, \FG{KapTransEF}(a), right column.  While the transmission at
$E_{\rm F}$ for the Co$_2$TiGe spacer shows large contributions near
$\vec k_\parallel=\bar{\Gamma}$ (corresponding to normal incidence),
these channels are almost blocked for Co$_2$TiSi (the dark blue area
around $\bar{\Gamma}$). In turn, Co$_2$TiSi favors transmission
channels far from $\bar{\Gamma}$, located along the diagonals of the
2D-BZ which appear very weak in Co$_2$TiGe.  Here we recall that the
Al/Co$_2$TiGe/Al (Co-Co) system 
has been found to be the only trilayer structure
for which ${\cal T}_\uparrow(E)$ passes $E_{\rm F}$ with a 
positive slope.
The redistribution of weights among the various transmission channels
in the 2D-BZ could explain the peculiar interface sensitivity
of ${\cal T}_\uparrow(E)$ and thus of $S_\uparrow(T)$, $S_{\rm
  eff}(T)$, and $S_{\rm spin}(T)$ evidenced above.

\begin{figure*}
  \centering
 \includegraphics[width=16cm]{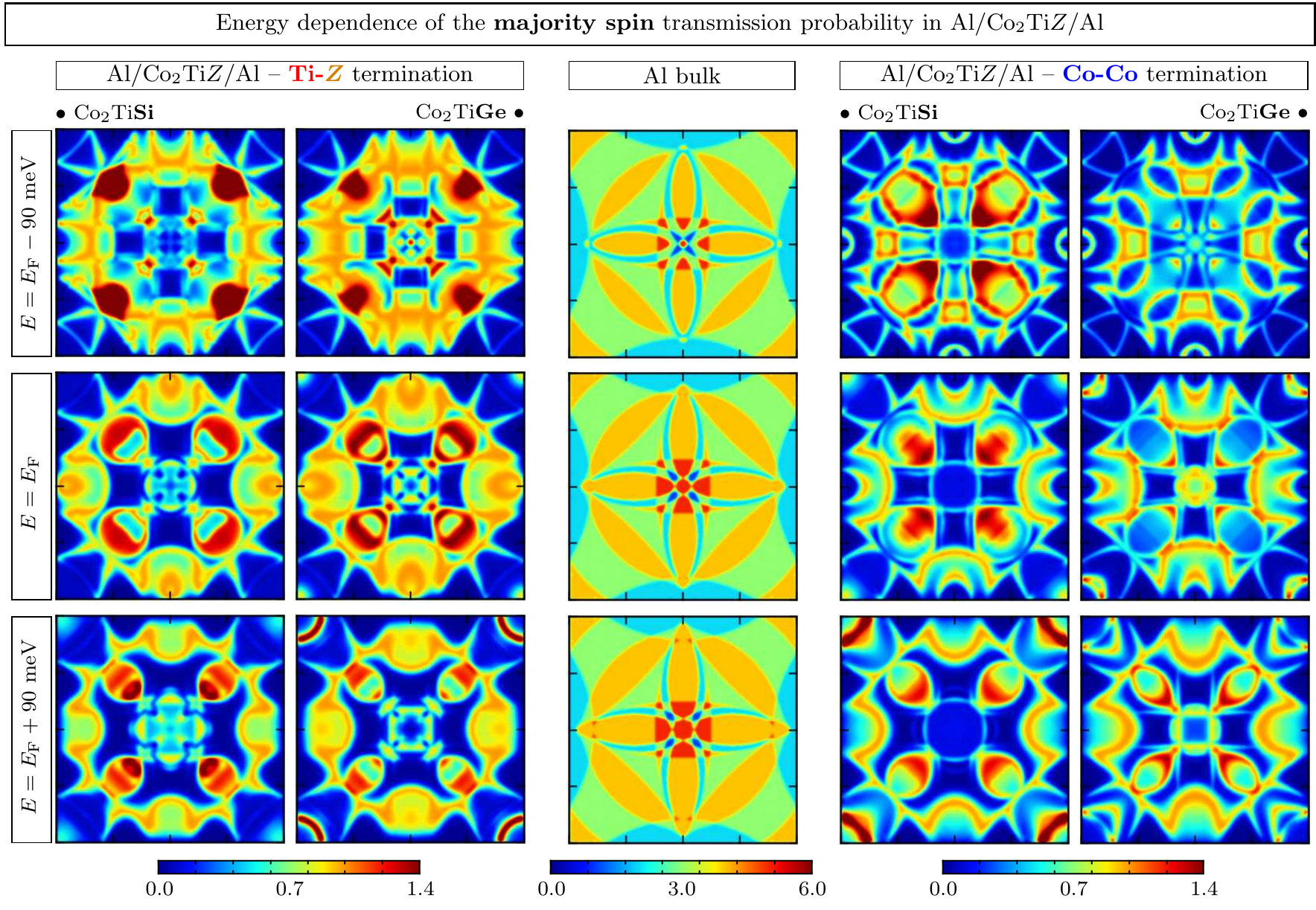}
 \caption{
   Contour plots of the majority spin $\vec
   k_\parallel$-resolved transmission probability in
   Al/Co$_2$Ti$Z$/Al for various energy arguments:
   $E=E_{\rm F}-90$~meV, $E=E_{\rm F}$, and 
   $E=E_{\rm F}+90$~meV, from top to bottom. 
   The left (right) columns correspond to
   the Ti-$Z$- (Co-Co-) terminated Al/Co$_2$Ti$Z$ interface.
   The middle column shows (on a different scale) the transmission 
   through unstrained Al bulk.
   The BZ is identical to that of \protect\FG{KapTransEF}.}
 \label{KapTransEVAR}
\end{figure*}

The final evidence must come from the {\em energy dependence} of the
various transmission channels. To make this point, we show in
Figure~\ref{KapTransEVAR} contour plots tracking the evolution of
${\cal T}_\uparrow(\vec k_\parallel,E)$ for several energy arguments,
$E_{\rm F}-90$~meV, $E_{\rm F}$, and $E_{\rm F}+90$~meV.  The layout
of this figure is similar to that of \FG{KapTransEF}
(Ti-$Z$-terminated systems on the left and Co-Co-terminated on the
right), but we have added the transmission maps of bulk Al leads. They
represent the incoming/receiving available transmission channels and
correspond to hypothetical transmission maps unaffected by the
electronic structure of the Heusler spacers and the Al/Heusler
interfaces.  The map for $E=E_{\rm F}$ is equivalent to the projected
Fermi surface of Al.

%FFFFFFFFFFFFFFFFFFFFFFFFFFFFFFFFFFFFFFFFFFFFFFFFFFFFFFFFFFFFFFFFFFFFFFFFFFFFF
%FFFFFFFFFFFFFFFFFFFFFFFFFFFFFFFFFFFFFFFFFFFFFFFFFFFFFFFFFFFFFFFFFFFFFFFFFFFFF
%FFFFFFF
%FFFFFFF
%FFFFFFFFFFFFFFFFFFFFFFFFFFFFFFFFFFFFFFFFFFFFFFFFFFFFFFFFFFFFFFFFFFFFFFFFFFFFF
%FFFFFFFFFFFFFFFFFFFFFFFFFFFFFFFFFFFFFFFFFFFFFFFFFFFFFFFFFFFFFFFFFFFFFFFFFFFFF

By showing the comparison to the transmission function of the Al
leads, we demonstrate that the absence of transmission near $\bar{\Gamma}$
for the Al/Co$_2$TiSi/Al (Co-Co) trilayer is {\em not} induced by the
leads, but rather an intrinsic materials property of this Heusler alloy.
Moreover, taking just Al bulk, one would expect the intensity and
width of the transmission around the zone center to increase. As can
be seen, the Al/Co$_2$TiSi/Al trilayer with Co-Co termination lacks
such transmission channels for {\em all} energy arguments.  
In turn, the amplitude of ${\cal T}_\uparrow(\vec k_\parallel,E)$
for the transmission channels along the 2D-BZ diagonals
is almost independent of energy. 
This implies that ${\cal T}_\uparrow(E)$ exhibits only a weak
energy dependence and thus $S_{\uparrow}$ is very small. 
This behavior is
in contrast to the Al/Co$_2$TiGe/Al system 
with Co-Co termination: Apart from
the presence of the additional transmission channel at $\bar{\Gamma}$, the
contributions from transmission channels along the 2D-BZ diagonals
tend to increase with energy. Thus, these
channels contribute with negative sign to $S_{\uparrow}$.
The Ti-$Z$ terminated systems, on the other hand, share qualitatively
similar features, with small quantitative differences, over a wide
energy range.

%SSSSSSSSSSSSSSSSSSSSSSSSSSSSSSSSSSSSSSSSSSSSSSSSSSSSSSSSSSSSSSSSSSSSSSSSSSSSSSS
%SSSSS
%SSSSS
\section{Summary}\label{SecFinal}

We have investigated the structural stability and the 
potential applicability as thermally driven spin injectors
of heterojunctions consisting of Al(001) and thin films of
the closely lattice-matched Heusler alloys Co$_2$TiSi 
and Co$_2$TiGe. Our most important findings can be summarized 
as follows: 

(i) The Al structure can be continued by Heusler alloys
terminated either by Co-Co or Ti-$Z$ ($Z=$ Si or Ge) planes; 
{\em ab initio} thermodynamic calculations predict that both 
terminations are stable against Al and Co$_2$Ti$Z$ separation.
Formation of Ti-$Z$-terminated interfaces requires, however,
non-equilibrium growth conditions.

(ii) As a consequence of a smaller equilibrium Co-Al bond length
than Ti-Al or Si(Ge)-Al, the structural relaxation occurring at 
the interface leads to a smaller separation between Al and the 
Heusler alloy spacer in the case of a Co-Co-terminated interface. 
This configuration appears to favor the Co-Al hybridization 
across the interface. 

(iii) The small lattice mismatch between the two components
(Al and Heusler alloy) results in a
sharp localization of the interface specific distortions, both
morphologic and in the electronic structure. Within $3-4$ MLs 
away from the interface the Heusler alloys accommodate an epitaxial,
tetragonally distorted structure with a preserved 
half-metallicity. The systems can be still regarded as
metal/half-metal heterojunctions, but the effective thickness 
of the half-metallic spacer is smaller than the 
geometric thickness of the Heusler film.

(iv) For the position of the Fermi level found in undoped samples, 
the effective and spin-dependent Seebeck coefficients 
are dominated by the majority spin carriers and 
significantly depend on the particular
spacer/termination combination. Combined with the 
growth conditions dependence of the formation of 
a specific stable interface, this establishes a direct connection 
between growth and spin-caloric properties control in 
these samples.

(v) The Ti-$Z$-terminated 
Heusler spacer layers exhibit qualitative similar features with slight
quantitative differences caused by varying the group IV
element $Z$. In both cases, the spin-dependent Seebeck
coefficient is predicted to be large, actually 
of the same size as the effective Seebeck coefficient. This
practically means that nearly the entire voltage generated
under a temperature gradient is converted into spin 
accumulation! 

(vi) For the Co-Co termination, the transport properties of the two
Heusler materials studied by us were quantitatively {\em and}
qualitatively different. In particular, we could show that the
Co-Co-terminated
Al/Co$_2$TiSi/Al heterojunction exhibits a much weaker
energy dependence of transmission and thus a very small (both
effective and spin-dependent) Seebeck coefficient.  For the
Co-Co-terminated
Al/Co$_2$TiGe/Al heterojunction, the obtainable
spin-dependent Seebeck coefficient is limited by the partial
cancellation of contributions from spin-up and spin-down electrons.

In summary, we conclude that the Al/Co$_2$Ti$Z$/Al trilayers
with Ti-$Z$ termination appear to be most suitable for achieving 
thermally driven spin injection of majority spin carriers. 
The actual growth of such systems has been shown to 
be attainable under out of equilibrium conditions.
In order to 
exploit the semiconductor-like electronic structure of the Heusler
alloys for minority spin injection, an additional fine-tuning of the
Fermi level position in these materials would be required. If the
Fermi energy could be brought close to the 
minority spin conduction band edge, e.g., by some way
of doping, it may be possible to obtain a material that displays large
$S_{\rm spin}$ due to the enhanced contribution of the minority spin
electrons.
Regardless of the type of carriers chosen, our calculations show that a
rather significant spin-dependent potential can be generated in 
the Al/Heusler/Al junctions under a thermal gradient provided that
the neglected scattering processes (electron-magnon and
electron-phonon) have a reduced deleterious character. The remaining
issue, not treated here, is the actual spin injection into a
semiconductor, such as Si, brought in contact with the trilayer.
The resulting heterostructure, Al/Heusler/Al/Si, would certainly 
require a careful choice of the thickness of the second Al layer,
which acts as buffer between the Heusler alloy and the semiconductor.
For a thin Al layer, the situation corresponds to a
ballistic spin injection similar to that investigated theoretically 
by Mavropoulos for the Fe/Si system,\cite{Mav08} 
and could be treated by a similar method as employed here.
Alternatively, the thickness of the Al can be even larger, though 
not exceeding the spin relaxation length. In this case,
explicit calculations can be carried out separately for 
the Al/Heusler/Al and Al/Si systems and their results combined
within the spin-charge coupling model developed by Scharf 
\ea\cite{SMAZF12} We certainly hope that such theoretical 
investigations will trigger corresponding experimental efforts.

%%%%%%%%%%%%%%%%%%%%%%%%%%%%%%%%%%%%%%%%%%%%%%%%%%%%%%%%%%%%%
%%%%%%%%%%%%%%%%%%%%%%%%%%%%%%%%%%%%%%%%%%%%%%%%%%%%%%%%%%%%%%%%%%%%%%%%%%

\acknowledgments

This work was supported by the German Research Foundation 
({\em Deutsche Forschungsgemeinschaft -- DFG}) 
within the Priority Program 1538 ''Spin Caloric Transport
(SpinCaT)''. 
The authors gratefully acknowledge the computing time granted by the 
John von Neumann Institute for Computing (NIC) and provided on the 
supercomputer JUROPA at J\"ulich Supercomputing Centre (JSC).
Additional computer facilities have been offered by the
Center for Computational Sciences and Simulation
(CCSS) at the University Duisburg-Essen. 
VP would also like to thank Dr.\ Phivos Mavropoulos for 
many enlightening discussions on 
the electronic transport formalism.

%%%%%%%%%%%%%%%%%%%%%%%%%%%%%%%%%%%%%%%%%%%%%%%%%%%%%%%%%%%%%%%%%%%%%%%%%%%
%%%%%%%%%%%%%%%%%%%%%%%%%%%%%%%%%%%%%%%%%%%%%%%%%%%%%%%%%%%%% LITERATURE

%%%%%%%%%%%%%%%%%%%%%%%%%%%%%%%%%%%%%%%%%%%%%%%%%%%%%%%%%%%%% LITERATURE
%%%%%%%%%%%%%%%%%%%%%%%%%%%%%%%%%%%%%%%%%%%%%%%%%%%%%%%%%%%%%%%%%%%%%%%%%%%


\begin{thebibliography}{49}
\bibitem[{Fol()}]{FoldingNote}
\bibinfo{note}{When using an 8-atom tetragonal unit cell of height
  $c=a_0\sqrt{2}$ the folded $Z$-point of the fcc BZ does not fall half-way
  between the $\Gamma$ and $Z$ points of the tetragonal BZ in
  Fig.~\ref{EvsKinEPI}. The same applies to the $X$-$R$ and $M$-$A$
  directions.}
\end{thebibliography}
\end{document}